\documentclass[journal]{IEEEtran}
\usepackage{amsmath,amsfonts,amsthm,amssymb,bm}
\usepackage{algorithmic}
\usepackage{array}
\usepackage[caption=false,font=normalsize,labelfont=sf,textfont=sf]{subfig}
\usepackage{textcomp}
\usepackage{stfloats}
\usepackage{url}
\usepackage{verbatim}
\usepackage{graphicx}
\usepackage{bbding}
\usepackage{multirow}
\usepackage{cite, bbm, xcolor}
\usepackage{refcount}
\usepackage{flafter}
\usepackage{makecell}
\usepackage{diagbox}
\usepackage{booktabs}
\usepackage{algorithm}

\theoremstyle{definition}

\usepackage{float}
\def\BibTeX{{\rm B\kern-.05em{\sc i\kern-.025em b}\kern-.08em
    T\kern-.1667em\lower.7ex\hbox{E}\kern-.125emX}}
    \expandafter\def\expandafter\normalsize\expandafter{%
    \normalsize%
    \setlength\abovedisplayskip{4pt}%
    \setlength\belowdisplayskip{4pt}%
    \setlength\abovedisplayshortskip{2pt}%
    \setlength\belowdisplayshortskip{2pt}%
}
\allowdisplaybreaks[4]
\usepackage{balance}

\begin{document}
\title{Multimodal LLM-Assisted Optimization for Pinching-Antennas-Enabled ISAC}
 \author{Deqiao Gan, Chongjun Ouyang, Cheng-Xiang Wang, and Zhiguo Ding\vspace{-10pt}
         \thanks{D. Gan and Z. Ding are with the School of Electrical and Electronic Engineering, Nanyang Technological University, Singapore.}
         \thanks{C. Ouyang is with the School of Electronic Engineering and Computer Science, Queen Mary University of London, London E1 4NS, U.K.}
         \thanks{C.-X. Wang is with the National Mobile Communications Research Laboratory, School of Information Science and Engineering, Southeast University, Nanjing.}
         }
\maketitle
\begin{abstract}
    The pinching-antenna systems (PASS) provide spatial reconfigurability through flexible pinching antenna (PA) placement along dielectric waveguides, which offers new opportunities for integrated sensing and communication (ISAC). However, the joint optimization of PA placement and digital beamforming leads to a highly nonconvex problem. To address this challenge, this paper proposes MAGIC-PASS, a multimodal large language model (LLM)-aware generative ISAC framework for PASS, which maximizes the sensing rate subject to minimum communication-rate requirements. MAGIC-PASS exploits multimodal observations of the wireless environment to jointly configure PA placement and digital beamforming. Specifically, the multimodal observations are encoded into continuous tokens and processed by an LLM to generate multiple candidate PASS configurations, each consisting of PA positions and digital beamformers. A diffusion model subsequently refines these candidates to improve generation stability. The feasible configuration with the highest sensing rate is then selected under the PA placement and transmit-power constraints. This generation-then-refinement procedure avoids a naive one-shot configuration and improves the reliability of the generated solutions. Numerical results demonstrate that MAGIC-PASS achieves a better sensing-communication tradeoff than conventional PASS, LLM-assisted multiple-input multiple-output (MIMO), and conventional MIMO schemes.
\end{abstract}


\section{Introduction}
Integrated sensing and communication (ISAC) enables wireless communication and sensing to share spectrum, radio-frequency hardware, and transmitted signals \cite{Liu2022ISAC,Zhang2021Overview,Liu2020JRC}. Beyond improving spectrum and hardware utilization, the coordinated design of the two functions enables communication-assisted sensing and sensing-assisted communication, which are important for networks requiring both reliable connectivity and environmental awareness. The benefits of ISAC mainly depend on the system's ability to direct energy toward communication users and sensing targets while reliably collecting the target echoes. To this end, multiple-input multiple-output (MIMO) techniques have been widely used to enable ISAC, which can adjust transmit and receive beamformers to strike a balanced sensing-communication performance \cite{wang2025hybrid}. However, the antenna locations and physical apertures in conventional MIMO arrays remain fixed after deployment. Consequently, only the signal weights can be adapted, whereas the underlying propagation geometry remains irreconfigurable. This spatial rigidity motivates development of spatially reconfigurable antennas that can adjust array geometry.


Fluid antenna system (FAS) and movable antenna (MA) are representative spatially reconfigurable antenna architectures \cite{Wong2021FAS,Zhu2024MA}. The FAS adjusts its effective radiating position by selecting among available ports or states, whereas the MA physically moves an antenna element within a prescribed region. Both architectures can exploit local spatial channel variations to improve the link quality. Their practical reconfiguration ranges, however, are generally limited by a finite aperture near the transceiver. Consequently, changing an antenna position within that aperture may have little effect on the free-space propagation distance to a remote user or target. It may also be insufficient to establish an unobstructed line-of-sight (LoS) link when signal blockage takes place. These limitations are particularly relevant at high frequencies, where large-scale path loss and blockage-induced attenuation can severely impair wireless links \cite{yuanwei2025pass}.

The pinching-antenna system (PASS), introduced by NTT DOCOMO, provides a means of extending the range of spatial reconfiguration \cite{2022NTTDOCOMO}. In PASS, a dielectric waveguide carries electromagnetic signals, while small dielectric elements placed at selected locations form pinching antennas (PAs) that couple energy between the waveguide and free space \cite{Ding2025PASS}. This architecture supports a flexible and scalable deployment strategy, which we refer to as pinching beamforming \cite{yuanwei2025pass}. A waveguide deployed across a wider service area can therefore bring a PA closer to a user or target, shortening the corresponding propagation distance and reducing its free-space path loss. When a waveguide location has an unobstructed path, PA placement can also improve LoS availability. Meanwhile, the PA positions determine both the free-space channels and the phases accumulated during in-waveguide propagation \cite{Wang2025Modeling}. 

Recent studies have investigated PA placement for PASS-enabled ISAC under different sensing objectives. In \cite{Zhang2025PASSISAC}, the PA positions were optimized to maximize target illumination power subject to a communication quality-of-service (QoS) requirement. A multi-waveguide architecture was studied in \cite{Li2026PASSISAC}, where the transmit PA positions and digital beamforming were jointly optimized to minimize the sensing Cramér--Rao bound. The communication--sensing rate region of a single-waveguide PASS serving one user and one target was characterized in \cite{Ouyang2026RateRegion}.
However, designing PASS for ISAC remains challenging. Specifically, the PA placement affects both the free-space and in-waveguide channels, which are coupled with digital beamforming to influence the performance of both sensing and communication. This coupling makes the resulting optimization problem nonconvex. Moreover, the position-dependent phase variations at each waveguide can cause the sensing and communication objectives to oscillate rapidly and create numerous local optima \cite{Xu2026LoSBlockage,Liu2026PASSTutorial}, which further complicates the joint design. Conventional alternating optimization (AO) and successive-approximation methods require iterative updates, and their performance depends strongly on initialization. Obtaining a high-quality PASS configuration requires multiple initializations, which can be computationally costly. Besides, once the channel realization or QoS change, the entire procedure must be re-performed, which is infeasible for PASS-ISAC deployments.

As an alternative, learning-assisted methods transfer part of the computational burden to offline training \cite{Guo2025Learning,Gao2026RLPASS}. A graph neural network was developed in \cite{Guo2025Learning} to determine PA positions and beamforming in a low-altitude ISAC network. The authors in \cite{Gao2026RLPASS} employed heterogeneous-graph reinforcement learning to optimize PA positions, time allocation, and transmit power under sensing-rate constraints. These methods reduce the online computational burden by learning a mapping from environmental information to resource allocation decisions. However, existing learning-based PASS designs mainly rely on the geometric positions of users and targets, which can be obtained through conventional positioning systems, such as GPS. Although such positional information characterizes part of the propagation environment and affects the path loss and phase shifts of wireless signals, it cannot fully capture the underlying propagation conditions. In particular, signal blockage, environmental changes, and historical channel state information (CSI) contain additional propagation information that cannot be inferred from user and target positions alone. Such information can be obtained from heterogeneous observations, including light detection and ranging (LiDAR) measurements and historical CSI records. Specifically, LiDAR measurements provide complementary information about scene geometry and blockage conditions, whereas historical CSI reflects the channel responses under previous PA configurations.
The position, LiDAR, blockage, and historical-CSI inputs provide information relevant to the user and target channels. For visual services, the encoded image size and transmission delay limitation specify the corresponding minimum user-rate requirement. These inputs inform the PASS configuration under both propagation conditions and communication-QoS requirements.
Therefore, these multimodal observations can provide a more comprehensive characterization of the wireless propagation environment and offer additional opportunities for optimizing PA placement and beamforming in PASS-enabled ISAC systems.


Despite these potential benefits, the exploitation of multimodal environmental information for PASS-enabled ISAC remains largely unexplored. A fundamental challenge lies in the heterogeneous nature of these observations, which differ in data formats, dimensions, and physical characteristics. Consequently, it is difficult to integrate them into a unified representation and extract the environmental features relevant to PA placement and beamforming optimization. Fortunately, large language models (LLMs) offer a possible approach to processing heterogeneous environmental information and facilitating wireless transmission design \cite{Xu2024LMM}. For example, LLMs have been investigated for millimeter-wave (mmWave) beam prediction \cite{wang2026beamagent}. The authors in \cite{Gan2025LLMPASS} incorporated visual observations, LiDAR measurements, and user locations into LLM-assisted codebook generation and beam selection. In \cite{Zheng2025M2BeamLLM}, multimodal observations from images, radar, LiDAR, and positioning systems were integrated into an LLM-based framework for mmWave beam prediction. Nevertheless, these studies primarily focus on beam prediction and selection, whereas the exploitation of LLMs for joint PA placement and digital beamforming in PASS-enabled ISAC remains unexplored.


Motivated by the above considerations, we propose Multimodal LLM-Aware Generative ISAC for PASS (MAGIC-PASS). The main contributions of this work are summarized as follows:
\begin{itemize}
    \item We investigate a PASS-enabled ISAC system that serves multiple communication users and senses a single target. Unlike conventional designs that primarily rely on user and target locations, we exploit multimodal environmental information, i.e., location and blockage from LiDAR measurements and historical CSI as well as visual observations, to facilitate PASS configurations. Based on this system model, we formulate a sensing-rate maximization problem that jointly optimizes transmit and receive PA placement and digital beamforming under communication-rate, PA placement, and transmit-power constraints.
    \item To address the resulting highly coupled and nonconvex optimization problem, we develop MAGIC-PASS, a multimodal LLM-assisted generative framework for adaptive PASS configuration. Specifically, we first develop a multimodal feature extraction and tokenization module to encode heterogeneous environmental observations and task requirements into continuous tokens. We then employ a pretrained LLM to process these tokens and generate multiple candidate PASS configurations, each of which specifies the transmit and receive PA placement and digital beamforming. To further improve the generated configurations, we introduce a conditional diffusion model to refine these candidates under the communication-rate, PA placement, and power constraints. Finally, we select the feasible configuration that achieves the highest sensing rate. Through this generation-then-refinement procedure, MAGIC-PASS exploits complementary environmental information to adapt PA placement and beamforming to varying propagation conditions and task requirements.
    \item We conduct extensive numerical simulations to evaluate the effectiveness of MAGIC-PASS. The results demonstrate that: i) MAGIC-PASS achieves substantial sensing-rate gains over conventional PASS, LLM-assisted MIMO, and conventional MIMO schemes while satisfying the communication-rate requirements; ii) the exploitation of multimodal environmental information improves the adaptability of PASS configurations across user--target geometries and blockage states compared with the single-modality variants; and iii) the proposed generation-then-refinement procedure improves the quality of PASS configurations compared with direct LLM-based generation.
\end{itemize}

The remainder of this paper is organized as follows.
Section~\ref{Section: System_Model} presents the system model and problem
formulation. Section~\ref{subsec:tokenization} introduces the multimodal feature extraction and tokenization modules of MAGIC-PASS architecture.
Section~\ref{subsec:candidate_configuration_generation} develops the LLM-assisted generation of PASS configuration candidate and its offline-training procedure.
Section~\ref{subsec:diffusion_refinement} presents the online LLM-assisted generative optimization procedure and its computational complexity.
Section~\ref{simulation_results} presents the numerical results, and
Section~\ref{Section: Conclusion} concludes the paper.

\subsubsection*{Notations}
Scalars, vectors, and matrices are denoted by plain, bold lowercase, and bold uppercase letters, respectively. The operators $(\cdot)^{\mathsf T}$ and $(\cdot)^{\mathsf H}$ denote transpose and Hermitian transpose. The symbols $|a|$, $\|\mathbf a\|_2$, and $\|\mathbf A\|_{\rm F}$ denote the magnitude of a scalar, the Euclidean norm of a vector, and the Frobenius norm of a matrix, respectively. The term $j\in\mathcal K_{\rm c}\setminus\{k\}$ means that $j \in \mathcal K_{\rm c}$ but $j \neq k$. The operator $\operatorname{vec}(\mathbf A)$ stacks the columns of $\mathbf A$, and $\mathbb E[\cdot]$ denotes expectation. For a positive integer $J$, $[J]\triangleq\{1,\ldots,J\}$. The notation $\mathcal{CN}(\boldsymbol\mu,\boldsymbol\Sigma)$ denotes a circularly symmetric complex Gaussian distribution with mean $\boldsymbol\mu$ and covariance $\boldsymbol\Sigma$. Finally, $[x]_+\triangleq\max\{x,0\}$, and $\mathcal O(\cdot)$ denotes asymptotic complexity.

\section{System Model}\label{Section: System_Model}
\begin{figure}[t]
    \centering
    \includegraphics[width=\columnwidth]{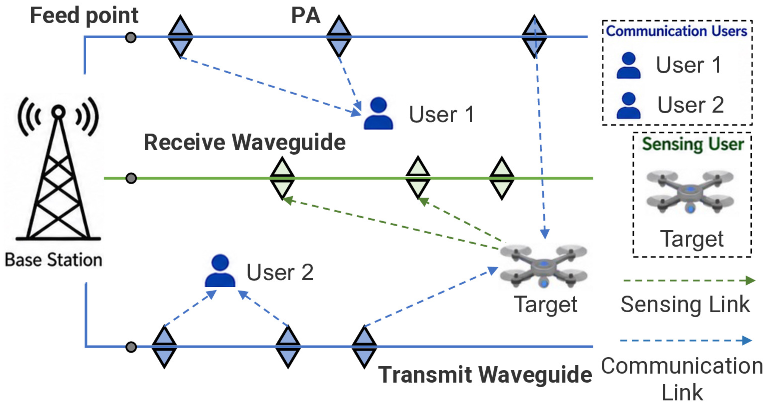}
    \caption{System model of the considered PASS-enabled ISAC system.}
    \label{system_model}
\end{figure}

Consider the PASS-enabled ISAC system in Fig.~\ref{system_model}, where a BS simultaneously serves $K_{\rm c}$ single-antenna communication users and senses one target. The BS employs $N_{\rm t}$ dielectric waveguides for transmission.  All waveguides are parallel to the $x$-axis.

The position of user $k\in\mathcal K_{\rm c}$ is $\mathbf u_k=[x_k,y_k,z_k]^{\mathsf T}$, and $\mathbf U\triangleq[\mathbf u_1,\ldots,\mathbf u_{K_{\rm c}}]$. The target's position available at the current configuration epoch is denoted by $\mathbf u_{\rm tar}=[x_{\rm tar},y_{\rm tar},z_{\rm tar}]^{\mathsf T}$. This position is supplied by a preceding localization or tracking procedure and is treated as fixed within one coherent processing interval. The sensing task considered here is therefore to estimate the target reflection coefficient rather than its position.

On each transmit waveguide, $M_{\rm t}$ PAs are activated, and the $m$-th PA along the $n$-th transmit waveguide is located at $\boldsymbol\phi_{{\rm t},m,n}=[x_{{\rm t},m,n},y_{{\rm t},n}, z_{{\rm t},n}]^{\mathsf T}$. The feed point of the $n$-th transmit waveguide is $\boldsymbol\phi_{{\rm t},0,n}=[x_{{\rm t},0,n},y_{{\rm t},n}, z_{{\rm t},n}]^{\mathsf T}$. The PA coordinates along the $x$-axis satisfy $x_{{\rm t},0,n}\leq x_{{\rm t},1,n}<x_{{\rm t},2,n}<\cdots<x_{{\rm t},M_{\rm t},n}\leq x_{{\rm t},0,n}+D_{\rm t}$, and adjacent transmit PAs are separated by at least a distance of $\Delta$. Define the position matrices of transmit PA along the $n$-th waveguide and transmit PA along all transmit waveguides by $\boldsymbol\Phi_{{\rm t},n}\triangleq [\boldsymbol\phi_{{\rm t},1,n},\ldots, \boldsymbol\phi_{{\rm t},M_{\rm t},n}]$ and $\boldsymbol\Phi_{\rm t}\triangleq [\boldsymbol\Phi_{{\rm t},1},\ldots, \boldsymbol\Phi_{{\rm t},N_{\rm t}}]$, respectively. 

Activating multiple receive PAs on a continuous dielectric waveguide may cause inter-antenna radiation via uplinks because a captured signal can propagate along the waveguide and be reradiated by another PA \cite{Wang2025Modeling}. Moreover, since one waveguide with one PA can provide the receive degree of freedom to observe the single target echo, the BS employs one dielectric receive waveguide along with one active PA for receiving. The position of the receive PA and feed point on this receive waveguide are $\boldsymbol\phi_{\rm r}=[x_{\rm r},y_{\rm r},z_{\rm r}]^{\mathsf T}$ and $\boldsymbol\phi_{{\rm r},0}=[x_{{\rm r},0},y_{\rm r},z_{\rm r}]^{\mathsf T}$, respectively, with $0\le x_{\rm r}-x_{{\rm r},0}\le D_{\rm r}$.

\subsection{Multimodal Environment and Task Information}
\label{subsec:multimodal_environment_model}

In the PASS-enabled ISAC system, the user and target positions determine the free-space link distances, the associated path loss and phase shifts. However, the geometric information of users and targets does not reveal the underlying channel conditions or potential obstructions. LiDAR measurements provide observations of the surrounding scene and its three-dimensional geometry, while the blockage derived from these observations supports the evaluation of link visibility in the free-space channel model. Historical CSI records previous channel responses and distinguishes propagation changes from PA placements. In visual-data services, the visual observation can provide the encoded image from camera frames, and its transmission delay limitation determines the required user rate. The task information specifies these rate requirements and power budgets. Accordingly, we consider six input groups: locations, visual observations, LiDAR measurements, historical CSI, blockage, and task requirements.

\begin{itemize}
    \item \emph{LiDAR measurements:} Let $t$ index the current PASS configuration epoch. A LiDAR sensor observing the BS service region provides the current point cloud $\mathbf L_{\rm RD}^{[t]}\in\mathbb R^{N_{\rm p}\times D_{\rm p}}$, which contains $N_{\rm p}$ returns with $D_{\rm p}$ features, including three-dimensional coordinates. After registration with the system coordinate frame, returns associated with a user or a target are used to refine their initial position estimates when such associations are available. The remaining scene geometry is used to update the blockage. Consequently, LiDAR affects the communication and sensing channels through the refined positions and link visibility.
    \item \emph{Locations:} At the configuration epoch $t$, the BS obtains initial user-position estimates from a preceding downlink positioning procedure. An initial target-position estimate is obtained from sensing echoes received in previous intervals. When the users or target can be identified in the current LiDAR measurement, the corresponding estimates are refined using their associated point-cloud returns. Denote the resulting user and target positions by $\mathbf U^{[t]}=[\mathbf u_1^{[t]},\ldots,\mathbf u_{K_{\rm c}}^{[t]}]$ and $\mathbf u_{\rm tar}^{[t]}$, respectively, and an initial estimate is retained if no associated LiDAR returns are available. These resulting positions $\{\mathbf U^{[t]},\mathbf u_{\rm tar}^{[t]}\}$ are treated as fixed in the current epoch.
    \item \emph{Blockage:} From the registered scene geometry of LiDAR measurements, a blockage mask is constructed to determine the visibility of a link between arbitrary specified locations. Specifically, after LiDAR measurements are registered to the system coordinate frame, the obstacle information is mapped to $N_{\rm blk}$ spatial cells of the serving region. The resulting vectorized blockage mask is expressed as $\mathbf B^{[t]}=\mathcal B_{\rm blk}(\mathbf L_{\rm RD}^{[t]})=[B_\varrho^{[t]}]_{\varrho\in [N_{\rm blk}]}\in\{0,1\}^{N_{\rm blk}}$, where $\mathcal B_{\rm blk}(\cdot)$ denotes the occupancy-mapping function. For the cell $\varrho\in[N_{\rm blk}]$, $B_\varrho^{[t]}=1$ indicates that the cell is occupied by an obstacle, and $B_\varrho^{[t]}=0$ indicates that it is unoccupied. The link is visible when every intersected cell is unoccupied, and it is blocked if the line segment intersects at least one occupied cell. Thus, the amplitude visibility coefficient based on the blockage mask is expressed as follows:
    \begin{equation}
        b(\mathbf d,\mathbf f;\mathbf B^{[t]}) = \prod_{\varrho\in\mathcal J(\mathbf d,\mathbf f)}(1-B_\varrho^{[t]}),
        \label{eq:link_visibility_from_mask}
    \end{equation}
    where $\mathcal J(\mathbf d,\mathbf f)$ denotes the cells intersected by the line segment connecting locations $\mathbf d$ and $\mathbf f$. Although $\mathbf B^{[t]}$ is constructed from LiDAR measurements, it retains only cell-level occupancy and discards the individual point coordinates and return features in $\mathbf L_{\rm RD}^{[t]}$. Therefore, it is necessary to retain LiDAR measurements to refine the user and target position estimates and to provide detailed scene geometry to the configuration generator. The blockage mask is also retained to provide the link-visibility coefficients used in the free-space channel model. 
    Moreover, the blockage mask remains fixed during the current configuration interval, but the visibility of a PA--user or PA--target link can change as the candidate PA position changes within each epoch.
    \item \emph{Visual observations:} The camera provides a sequence of frames $\mathbf O^{[t]}\triangleq \{\mathbf O^{[t']}\}_{t'=t-\eta_{\rm o}+1}^{t}$, where $\eta_{\rm o}$ is the number of available frames. Each frame $\mathbf O^{[t']}\in \mathbb R^{H_{\rm o}\times W_{\rm o}\times C_{\rm o}}$ has height $H_{\rm o}$, width $W_{\rm o}$, and $C_{\rm o}$ channels. The frames describe the observed image content delivered through the downlink data streams. Specifically, in a visual-data service, the image content scheduled for the user $k$ is encoded into a payload of $D_k^{[t]}$ bits. To deliver this payload within transmission delay $T_k^{[t]}$ over bandwidth $f_{\rm bw}$, the required minimum user rate is $(R_k^{\min})^{[t]}=D_k^{[t]}/(f_{\rm bw}T_k^{[t]})$. Thus, visual observations can be used to determine the minimum user rate requirements.
    \item \emph{Task requirement:} We collect the minimum user-rate requirements from visual observations and the transmit-power budget in the task vector $\boldsymbol\tau^{[t]}\triangleq[(R_1^{\min})^{[t]},\ldots,(R_{K_{\rm c}}^{\min})^{[t]},P_{\max}^{[t]}]^{\mathsf T}$. When an image payload is delivered to the user $k$, its encoded size and transmission delay limitation determine the corresponding $(R_k^{\min})^{[t]}$. Although the task requirement does not change the propagation channels for a fixed PASS configuration, it can determine QoS requirements during the wireless transmission.
    \item \emph{Historical CSI:} The BS obtains effective downlink CSI from pilot measurements and user feedback. Since the measured channel depends on the PA positions used during acquisition, each CSI record is stored together with PA positions. The available historical CSI can be expressed as $\mathbf C^{[t]} \triangleq \{\mathbf H^{[t']}, \boldsymbol\Phi_{\rm t}^{[t']},\boldsymbol\phi_{\rm r}^{[t']}\}_{t'=t-\eta_{\rm CSI}+1}^{t}$,
    where $\eta_{\rm CSI}$ is the number of records and $\mathbf H^{[t']} = [\widetilde{\mathbf h}_1^{[t']},\ldots,\widetilde{\mathbf h}_{K_{\rm c}}^{[t']}]^{\mathsf H} \in\mathbb C^{K_{\rm c}\times N_{\rm t}}$ contains the effective user channels measured under $\boldsymbol\Phi_{\rm t}^{[t']}$. Since each channel measurement is associated with its acquisition configuration, these records provide the configuration generation with observed channel responses at previously used PA positions. They supplement the current scene observations with historical channel information when generating PA positions and beamforming for the next configuration.
\end{itemize}

Therefore, the six input categories of information are summarized as follows:
\begin{equation}
    \label{eq:system_multimodal_input}
    \mathcal E^{[t]}\triangleq \{\mathbf U^{[t]},\mathbf u_{\rm tar}^{[t]},
    \mathbf O^{[t]},\mathbf L_{\rm RD}^{[t]},\mathbf C^{[t]},\mathbf B^{[t]},
    \boldsymbol\tau^{[t]}\}.
\end{equation}

\subsection{Communication Model}
Radio-channel measurements with antennas have highlighted the coupling between antenna settings and wireless propagation~\cite{Wang2026ReconfigurableChannels}. More comprehensive 6G channel models address a wider range of propagation conditions, while continuous-space models examine the dependence of channels on antenna placement \cite{Wang2025PervasiveChannel,Wang2026ContinuousSpace}.
For a tractable PASS configuration design, we adopt an LoS-dominant PASS channel model \cite{Ding2025PASS,Wang2025Modeling}. Let $\lambda$ and $\lambda_{\rm g}=\lambda/n_{\rm eff}$ denote the free-space and guided wavelengths, respectively, where $n_{\rm eff}$ is the effective refractive index. With $d_{{\rm t},m,n}\triangleq\|\boldsymbol\phi_{{\rm t},m,n}- \boldsymbol\phi_{{\rm t},0,n}\|_2$, the in-waveguide response of the $m$-th PA on the $n$-th transmit waveguide is expressed as follows:
\begin{equation}
    g_{\rm t}(\boldsymbol\phi_{{\rm t},m,n})
    =\frac{1}{\sqrt{M_{\rm t}}} e^{-\mathrm j\frac{2\pi}{\lambda_{\rm g}}d_{{\rm t},m,n}} 10^{-\frac{\kappa}{20}d_{{\rm t},m,n}},
\label{inwaveguidechannel}
\end{equation}
where $\kappa$ is the waveguide attenuation coefficient in dB/m \cite{yeh2008kappa}. The factor $1/\sqrt{M_{\rm t}}$ models equal power splitting among the PAs on each transmit waveguide. The PA placement changes both the corresponding free-space propagation distance and the accumulated in-waveguide phase. The transmit-side in-waveguide channel is denoted as follows:
\begin{equation} 
    \mathbf g_{{\rm t},n}\triangleq [g_{\rm t}(\boldsymbol\phi_{{\rm t},1,n}),\ldots, g_{\rm t}(\boldsymbol\phi_{{\rm t},M_{\rm t},n})]^{\mathsf T},
\label{transmit_waveguide_matrix}
\end{equation}
where $\mathbf G_{\rm t}(\boldsymbol\Phi_{\rm t})\triangleq \operatorname{blkdiag}(\mathbf g_{{\rm t},1},\ldots, \mathbf g_{{\rm t},N_{\rm t}}) \in\mathbb C^{M_{\rm t}N_{\rm t}\times N_{\rm t}}$ denotes the overall transmit-side in-waveguide channel matrix.


The associated free-space channel is expressed as $h(\mathbf d,\mathbf f;\mathbf B) =b(\mathbf d,\mathbf f;\mathbf B)\sqrt{\eta}\, \frac{e^{-\mathrm j k_0\|\mathbf d-\mathbf f\|_2}}{\|\mathbf d-\mathbf f\|_2}$,
where $\eta=c^2/(16\pi^2f_{\rm c}^2)$, $c$ is the speed of light, $f_{\rm c}$ is the carrier frequency, and $k_0=2\pi/\lambda$.
The channel vector from all transmit PAs to the user $k$ is given by
\begin{equation}
    \label{communication_free_space_vector}
    \mathbf h_{{\rm c},k} \triangleq [h(\mathbf u_k,\boldsymbol\phi_{{\rm t},1,1};\mathbf B),\ldots, h(\mathbf u_k,\boldsymbol\phi_{{\rm t},M_{\rm t},N_{\rm t}};\mathbf B)]^{\mathsf T}.
\end{equation}

The effective channel from the transmit waveguides is $\widetilde{\mathbf h}_k^{\mathsf H}\triangleq \mathbf h_{{\rm c},k}^{\mathsf H} \mathbf G_{\rm t}(\boldsymbol\Phi_{\rm t}) \in\mathbb C^{1\times N_{\rm t}}$.
Let $\mathbf s_k\in\mathbb C^{L\times1}$ be the length-$L$ data block for user $k$ and $\mathbf S\triangleq[\mathbf s_1,\ldots,\mathbf s_{K_{\rm c}}]^{\mathsf T} \in\mathbb C^{K_{\rm c}\times L}$. We adopt the standard block-normalization $\mathbf S\mathbf S^{\mathsf H}=L\mathbf I_{K_{\rm c}}$.
The digital beamforming matrix is $\mathbf W=[\mathbf w_1,\ldots,\mathbf w_{K_{\rm c}}] \in\mathbb C^{N_{\rm t}\times K_{\rm c}}$. The signal received by user $k$ is given by
\begin{equation}
    \mathbf y_k^{\mathsf T} =\widetilde{\mathbf h}_k^{\mathsf H}\mathbf w_k\mathbf s_k^{\mathsf T} +\sum_{j\in\mathcal K_{\rm c}\setminus\{k\}} \widetilde{\mathbf h}_k^{\mathsf H}\mathbf w_j\mathbf s_j^{\mathsf T} +\mathbf n_k^{\mathsf T},
    \label{receivedsignal}
\end{equation}
where $\mathbf n_k\sim\mathcal{CN}(\mathbf0,\sigma_{\rm c}^2\mathbf I_L)$.
The corresponding signal-to-interference-plus-noise ratio (SINR) is given by
\begin{equation}
    \gamma_k =\frac{|\widetilde{\mathbf h}_k^{\mathsf H}\mathbf w_k|^2}{\sum_{j\in\mathcal K_{\rm c}\setminus\{k\}}|\widetilde{\mathbf h}_k^{\mathsf H}\mathbf w_j|^2+\sigma_{\rm c}^2},
    \label{sinr}
\end{equation}
and the achievable rate can be expressed as follows:
\begin{equation}
    R_k(\mathbf U,\mathbf B,\boldsymbol\Phi_{\rm t},\mathbf W)=\log_2(1+\gamma_k).
    \label{communication_rate}
\end{equation}

\subsection{Sensing Model}

The communication waveform is reused for sensing \cite{Liu2020JRC,Liu2022ISAC}. With $d_{\rm r}\triangleq\|\boldsymbol\phi_{\rm r}-\boldsymbol\phi_{{\rm r},0}\|_2$, the channel from the receive PA to its feed point is expressed as follows:
\begin{equation}
    g_{\rm r}(\boldsymbol\phi_{\rm r})=e^{-\mathrm j\frac{2\pi}{\lambda_{\rm g}}d_{\rm r}}10^{-\frac{\kappa}{20}d_{\rm r}},
    \label{inwaveguidechannelreceive}
\end{equation}
where both $g_{\rm r}(\boldsymbol\phi_{\rm r})$ and $g_{\rm t}(\boldsymbol\phi_{{\rm t},m,n})$ in~\eqref{inwaveguidechannel} use $10^{-\kappa d/20}$ for amplitude attenuation when $\kappa$ is specified in dB/m. The factor $1/\sqrt{M_{\rm t}}$ appears only in $g_{\rm t}(\boldsymbol\phi_{{\rm t},m,n})$ because the power injected into a transmit waveguide is divided among its $M_{\rm t}$ active PAs. No corresponding splitting factor is required for the receive waveguide, which has one active PA.
The free-space transmit channel toward the target can be expressed as follows:
\begin{equation}
    \mathbf h_{\rm t}\triangleq [h(\mathbf u_{\rm tar},\boldsymbol\phi_{{\rm t},11};\mathbf B),\ldots,h(\mathbf u_{\rm tar},\boldsymbol\phi_{{\rm t},M_{\rm t}N_{\rm t}};\mathbf B)]^{\mathsf T},
    \label{freespacechanneltransmit}
\end{equation}
and its effective channel from the transmit waveguides is $\widetilde{\mathbf h}_{\rm t}^{\mathsf H}\triangleq \mathbf h_{\rm t}^{\mathsf H}\mathbf G_{\rm t}(\boldsymbol\Phi_{\rm t}) \in\mathbb C^{1\times N_{\rm t}}$. The effective target-to-receiver channel is given by
\begin{equation}
    \widetilde h_{\rm r} \triangleq g_{\rm r}(\boldsymbol\phi_{\rm r}) h(\boldsymbol\phi_{\rm r},\mathbf u_{\rm tar};\mathbf B).
    \label{freespacechannelreceive}
\end{equation}

Following the information-theoretic sensing model in \cite{Ouyang2026RateRegion}, let $\beta\sim\mathcal{CN}(0,\alpha_{\rm s})$ denote the target reflection coefficient, which remains constant over $L$ sensing snapshots, where $\alpha_{\rm s}$ denotes its average power. The received echo vector is given by
\begin{equation}
    \mathbf y_{\rm r}
    =\beta\widetilde h_{\rm r}(\widetilde{\mathbf h}_{\rm t}^{\mathsf H}\mathbf W\mathbf S)^{\mathsf T}+\mathbf n_{\rm s}=\beta\mathbf a+\mathbf n_{\rm s},
    \label{overallreceivedsensing}
\end{equation} 
where $\mathbf a\triangleq \widetilde h_{\rm r}(\widetilde{\mathbf h}_{\rm t}^{\mathsf H}\mathbf W\mathbf S)^{\mathsf T} \in\mathbb C^{L\times1}$ and $\mathbf n_{\rm s}\sim \mathcal{CN}(\mathbf0,\sigma_{\rm s}^2\mathbf I_L)$.
Conditioned on the transmitted block $\mathbf S$, the sensing mutual information is expressed as follows:
\begin{equation}
    I(\beta;\mathbf y_{\rm r}\!\mid\!\mathbf S) \!=\!\log_2\det\!(\mathbf I_L\!+\!\frac{\alpha_{\rm s}}{\sigma_{\rm s}^2} \mathbf a\mathbf a^{\mathsf H}) \!=\!\log_2(1\!+\!\frac{\alpha_{\rm s}}{\sigma_{\rm s}^2}\|\mathbf a\|_2^2).
\label{sensing_mutual_information}
\end{equation}

Accordingly, the achievable sensing rate per snapshot is given by
\begin{equation}
    \begin{aligned}
        &R_{\rm r}(\boldsymbol\Phi_{\rm t},\mathbf u_{\rm tar},\boldsymbol\phi_{\rm r},\mathbf B,\mathbf W) \triangleq \frac{1}{L}I(\beta;\mathbf y_{\rm r}\mid\mathbf S)\\
        &=\frac{1}{L}\log_2(1+\frac{L\alpha_{\rm s}}{\sigma_{\rm s}^2}|\widetilde h_{\rm r}|^2 \|\widetilde{\mathbf h}_{\rm t}^{\mathsf H}\mathbf W\|_2^2),
    \end{aligned}
\label{achievable sensing rate}
\end{equation}
where the second equality follows from $\mathbf S\mathbf S^{\mathsf H}=L\mathbf I_{K_{\rm c}}$.

\subsection{Problem Formulation}

For a given environment state $(\mathbf U,\mathbf u_{\rm tar},\mathbf B)$, we jointly optimize transmit-PA positions $\boldsymbol\Phi_{\rm t}$, receive-PA position $\boldsymbol\phi_{\rm r}$, and digital beamforming matrix $\mathbf W$. The sensing-rate maximization problem is formulated as follows:
\begin{subequations}
    \label{problem:P1}
    \begin{align}
        & \max_{\boldsymbol\Phi_{\rm t},\boldsymbol\phi_{\rm r},\mathbf W} \quad  R_{\rm r}(\boldsymbol\Phi_{\rm t},\mathbf u_{\rm tar}, \boldsymbol\phi_{\rm r},\mathbf B,\mathbf W)
        \label{max_sensing_rate}\\
        \mathrm{s.t.}\quad & R_k(\mathbf U,\mathbf B,\boldsymbol\Phi_{\rm t},\mathbf W) \ge R_k^{\min}, \ \forall k\in\mathcal K_{\rm c},\label{min_rate_constraint}\\
        & 0\le x_{{\rm t},m,n}-x_{{\rm t},0,n}\le D_{\rm t}, \ \forall m\in\mathcal M_{\rm t},\ n\in\mathcal N_{\rm t},
        \label{transmit_position_constraint}\\
        & 0\le x_{\rm r}-x_{{\rm r},0}\le D_{\rm r},
        \label{receive_position_constraint}\\
        & x_{{\rm t},m,n}\!-\!x_{{\rm t},(m-1),n}\ge\Delta, \ \forall m\in\mathcal M_{\rm t}\!\setminus\!\{1\},\ n\in\mathcal N_{\rm t},
        \label{transmit_spacing_constraint}\\
        &\|\mathbf W\|_{\rm F}^2\le P_{\max}.
        \label{power_constraint}
    \end{align}
\end{subequations}

Here, $R_k^{\min}$ is the minimum rate required by user $k$, and $P_{\max}$ is the BS transmit-power budget. Constraint~\eqref{min_rate_constraint} enforces the communication QoS, where communication-rate constraints are evaluated using the Gaussian-input achievable rate in~\eqref{communication_rate}; constraints \eqref{transmit_position_constraint} and \eqref{receive_position_constraint} restrict the PAs to their movement ranges; \eqref{transmit_spacing_constraint} limits mutual coupling between adjacent transmit PAs; and \eqref{power_constraint} imposes the transmit-power budget.


The transmit PA positions influence the free-space and in-waveguide responses through distance-dependent phase terms. Because the signals radiated by different PAs combine coherently, these phase terms make the sensing rate oscillatory and nonconcave in the transmit PA positions. The beamforming matrix $\mathbf W$ further affects the sensing rate and the multiuser-interference-coupled QoS constraints, while the receive PA position changes the target-echo strength through free-space path loss and in-waveguide attenuation. Consequently, problem~\eqref{problem:P1} is nonconvex. Moreover, when the environment or service requirements change, conventional iterative methods must repeatedly update $(\boldsymbol\Phi_{\rm t},\boldsymbol\phi_{\rm r},\mathbf W)$ to solve this nonconvex problem. These difficulties motivate the proposed LLM-assisted generative framework for exploring high-quality PASS configurations efficiently.

\section{Multimodal Feature Extraction and Tokenization}
\label{subsec:tokenization}

\begin{figure}[t]
    \centering
    \includegraphics[width=0.9\columnwidth]{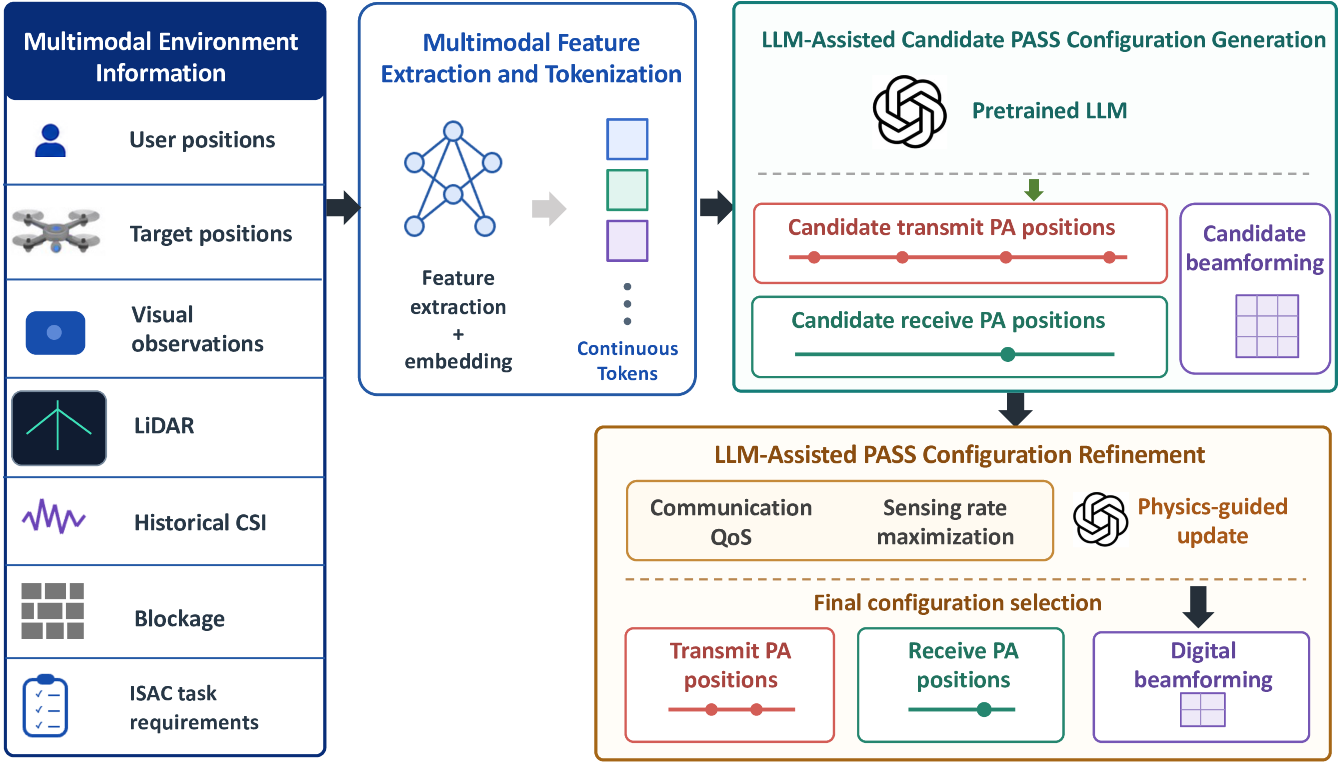}
    \caption{Illustration of the proposed MAGIC-PASS framework.}
    \label{MAGIC_PASS framework}
\end{figure}
To solve this nonconvex problem \eqref{problem:P1}, we now develop the MAGIC-PASS architecture to determine the transmit and receive PA positions, and digital beamforming matrix from six categories of inputs defined in Section~\ref{subsec:multimodal_environment_model}. Since the LLM can not process the heterogeneous multimodal inputs directly, the architecture first converts the six input categories into continuous tokens for a pretrained LLM, as illustrated in Fig.~\ref{MAGIC_PASS framework}. The resulting tokens are utilized to generate candidate PASS configurations, which are then refined and evaluated according to their sensing rates and communication-QoS requirements, described in the following Sections~\ref{subsec:candidate_configuration_generation} and~\ref{subsec:diffusion_refinement}.

For each information category $\mathcal{\vartheta} \in \{\mathrm{loc,vis,lid,CSI,blk,task}\}$, the encoder $\operatorname{Enc}_{\vartheta}(\cdot)$ extracts features from $\mathcal E^{[t]}$, and projection function $\operatorname{Proj}_{\vartheta}(\cdot)$ maps them to $D_{\rm M}$-dimensional continuous tokens. A trainable modality embedding $\boldsymbol\epsilon_{\vartheta}^{\rm mod}$ is added to identify the source of each token. Moreover, the same modality embedding is added to all tokens produced from a given input category. The six categories of encoders and projection functions are summarized in Table~\ref{tab:modality_modules}.

\begin{table}[t]
\centering
\caption{Encoder and projection for the six categories}
\label{tab:modality_modules}
\footnotesize
\setlength{\tabcolsep}{2.9pt}
\begin{tabular}{@{}lccc@{}}
\toprule
Input category & Encoder & Projection &
\shortstack{Modality embedding} \\
\midrule
Location
& $\operatorname{Enc}_{\rm loc}(\cdot)$
& $\operatorname{Proj}_{\rm loc}(\cdot)$
& $\boldsymbol\epsilon_{\rm loc}^{\rm mod}$ \\
Visual observation
& $\operatorname{Enc}_{\rm vis}(\cdot)$
& $\operatorname{Proj}_{\rm vis}(\cdot)$
& $\boldsymbol\epsilon_{\rm vis}^{\rm mod}$ \\
LiDAR
& $\operatorname{Enc}_{\rm lid}(\cdot)$
& $\operatorname{Proj}_{\rm lid}(\cdot)$
& $\boldsymbol\epsilon_{\rm lid}^{\rm mod}$ \\
Historical CSI
& $\operatorname{Enc}_{\rm CSI}(\cdot)$
& $\operatorname{Proj}_{\rm CSI}(\cdot)$
& $\boldsymbol\epsilon_{\rm CSI}^{\rm mod}$ \\
Blockage
& $\operatorname{Enc}_{\rm blk}(\cdot)$
& $\operatorname{Proj}_{\rm blk}(\cdot)$
& $\boldsymbol\epsilon_{\rm blk}^{\rm mod}$ \\
Task requirement
& $\operatorname{Enc}_{\rm task}(\cdot)$
& $\operatorname{Proj}_{\rm task}(\cdot)$
& $\boldsymbol\epsilon_{\rm task}^{\rm mod}$ \\
\bottomrule
\end{tabular}
\end{table}

The LiDAR measurement in the epoch $t$ is processed as follows:
\begin{equation}
    \label{eq:lidar_tokens}
    \mathbf Z_{\rm lid}^{\rm tok}
    = \operatorname{Proj}_{\rm lid}\!(\operatorname{Enc}_{\rm lid}(\mathbf L_{\rm RD}^{[t]}))+\boldsymbol\epsilon_{\rm lid}^{\rm mod},
\end{equation}
where $\mathbf Z_{\rm lid}^{\rm tok}\in\mathbb R^{N_{\rm lid}\times D_{\rm M}}$. A point-cloud encoder such as PointNet can be used for $\operatorname{Enc}_{\rm lid}$ \cite{Qi2017PointNet}. The resulting tokens describe the three-dimensional geometry of the surrounding objects. No temporal embedding is added because only the current LiDAR measurement is used.

The location tokens are constructed as follows:
\begin{subequations}
    \label{eq:location_tokens}
    \begin{align}
        \mathbf z_{{\rm usr},k}^{\rm tok}
        &=\operatorname{Proj}_{\rm loc}\!(\operatorname{Enc}_{\rm loc}(\mathbf u_k^{[t]}))+\boldsymbol\epsilon_{\rm usr}+\boldsymbol\epsilon_{{\rm usr},k}^{\rm idx}+\boldsymbol\epsilon_{\rm loc}^{\rm mod},\\
        \mathbf z_{\rm tar}^{\rm tok}
        &=\operatorname{Proj}_{\rm loc}\!(\operatorname{Enc}_{\rm loc}(\mathbf u_{\rm tar}^{[t]}))+\boldsymbol\epsilon_{\rm tar}+\boldsymbol\epsilon_{\rm loc}^{\rm mod},
    \end{align}
\end{subequations}
where the user-index embedding parameter $\boldsymbol\epsilon_{{\rm usr},k}^{\rm idx}$ preserves user identity in continuous tokens, and $\boldsymbol\epsilon_{\rm usr}$ and $\boldsymbol\epsilon_{\rm tar}$ distinguish the two entity types. A target index embedding parameter is unnecessary because only one target is considered. The location-token matrix is $\mathbf Z_{\rm loc}^{\rm tok}=\left[(\mathbf z_{{\rm usr},1}^{\rm tok})^{\mathsf T};\ldots;(\mathbf z_{{\rm usr},K_{\rm c}}^{\rm tok})^{\mathsf T};(\mathbf z_{\rm tar}^{\rm tok})^{\mathsf T}\right]\in\mathbb R^{(K_{\rm c}+1)\times D_{\rm M}}$.
The visual sequence is expressed as follows:
\begin{equation}
    \label{eq:visual_tokens}
    \mathbf Z_{\rm vis}^{\rm tok} = \underset{t'=t-\eta_{\rm o}\!+\!1\!,\ldots\!,t}{\operatorname{Stack}}[\operatorname{Proj}_{\rm vis}\!(\operatorname{Enc}_{\rm vis}(\mathbf O^{[t']}))+\boldsymbol\epsilon_{\rm vis}^{\rm mod}+\boldsymbol\epsilon_{t'}^{\rm time}],
\end{equation}
where $\mathbf Z_{\rm vis}^{\rm tok} \in\mathbb R^{N_{\rm vis}\times D_{\rm M}}$ and $N_{\rm vis}$ is the total number of tokens extracted from the $\eta_{\rm o}$ frames. The visual encoder may be implemented by a pretrained vision transformer \cite{Dosovitskiy2021ViT}. The temporal embedding $\boldsymbol\epsilon_{t'}^{\rm time}$ retains the order of the observations, thereby preserving the order of the available image frames.

The blockage tokens retain the visibility information used by the channel model, whereas the task token identifies the QoS and power constraints of the current instance of \eqref{problem:P1}. The blockage and task tokens are mapped as follows:
\begin{align}
    \mathbf Z_{\rm blk}^{\rm tok} & = \operatorname{Proj}_{\rm blk}\!(\operatorname{Enc}_{\rm blk}(\mathbf B^{[t]}))+\boldsymbol\epsilon_{\rm blk}^{\rm mod}\in\mathbb R^{N_{\rm blk}\times D_{\rm M}},
    \label{eq:blockage_tokens}\\
    \mathbf Z_{\rm task}^{\rm tok} & = \operatorname{Proj}_{\rm task}\!(\operatorname{Enc}_{\rm task}(\boldsymbol\tau^{[t]}))+\boldsymbol\epsilon_{\rm task}^{\rm mod}\in\mathbb R^{1\times D_{\rm M}}.\label{eq:task_tokens}
\end{align}

The CSI history includes the PA configuration used to acquire each channel sample. This information is necessary because a channel change can be caused by either environmental evolution or PA repositioning. For each epoch $t'$, define the real-valued CSI feature vector by
\begin{equation}
    \begin{aligned}
        \mathbf c^{[t']}\triangleq [&\operatorname{vec}^{\mathsf T}(\Re\{\mathbf H^{[t']}\}),\operatorname{vec}^{\mathsf T}(\Im\{\mathbf H^{[t']}\}),\\
        & \operatorname{vec}^{\mathsf T}(\boldsymbol\Phi_{\rm t}^{[t']}),(\boldsymbol\phi_{\rm r}^{[t']})^{\mathsf T}]^{\mathsf T} \in\mathbb R^{D_{\rm CSI}},
    \end{aligned}
\label{eq:csi_feature_vector}
\end{equation}
where $D_{\rm CSI}=2K_{\rm c}N_{\rm t}+3M_{\rm t}N_{\rm t}+3$. The configuration-aware CSI tokens are encoded as follows:
\begin{equation}
    \label{eq:csi_tokens}
    \mathbf Z_{\rm CSI}^{\rm tok} \!=\!\operatorname{Stack}(\{\operatorname{Proj}_{\rm CSI}\!(\operatorname{Enc}_{\rm CSI}(\mathbf c^{[t']}))\!+\!\boldsymbol\epsilon_{\rm CSI}^{\rm mod}\!+\!\boldsymbol\epsilon_{t'}^{\rm time}\}_{t'=t\!-\!\eta_{\rm CSI}\!+\!1}^{t}),
\end{equation}
where $\mathbf Z_{\rm CSI}^{\rm tok}\in\mathbb R^{\eta_{\rm CSI}\times D_{\rm M}}$ and $t'\in \{t-\eta_{\rm CSI}+1,t\}$. The channel components and PA coordinates are normalized using their respective training-set statistics before forming $\mathbf c^{[t']}$.


The six token sequences are concatenated via the stacking function to form the unified input-token sequence, given by
\begin{equation}
    \mathbf Z_{\mathcal E}^{\rm in} =\left[\mathbf Z_{\rm loc}^{\rm tok};\mathbf Z_{\rm vis}^{\rm tok}; \mathbf Z_{\rm lid}^{\rm tok}; \mathbf Z_{\rm CSI}^{\rm tok}; \mathbf Z_{\rm blk}^{\rm tok}; \mathbf Z_{\rm task}^{\rm tok}\right] \!\in \! \mathbb R^{N_{\mathcal E}\!\times\! D_{\rm M}},
    \label{eq:unified_multimodal_tokens}
\end{equation}
where $N_{\mathcal E}=(K_{\rm c}+1)+N_{\rm vis}+N_{\rm lid}+\eta_{\rm CSI}+N_{\rm blk}+1$.
The two occurrences of one correspond to the single target-location token and the task token, respectively. The resulting sequence is processed by the pretrained LLM backbone as follows:
\begin{equation}
    \label{eq:multimodal_llm_output}
    \mathbf Z_{\mathcal E}^{\rm out}=\operatorname{LLM}_{\boldsymbol\Theta_0}(\mathbf Z_{\mathcal E}^{\rm in})=\left[\mathbf Z_{\rm loc}^{\rm out};\mathbf Z_{\rm vis}^{\rm out};\mathbf Z_{\rm lid}^{\rm out};\mathbf Z_{\rm CSI}^{\rm out};\mathbf Z_{\rm blk}^{\rm out};\mathbf Z_{\rm task}^{\rm out}\right],
\end{equation}
where $\operatorname{LLM}_{\boldsymbol\Theta_0}(\cdot)$ denotes the pretrained LLM function with frozen LLM parameters $\boldsymbol\Theta_0$. The trainable encoders, projections and embedding parameters are learned during offline training based on the pretrained LLM and fixed during online optimization.

\section{LLM-Assisted Candidate PASS Configuration Generation}
\label{subsec:candidate_configuration_generation}

To deal with this nonconvex problem \eqref{problem:P1}, we propose the LLM-assisted candidate PASS configuration generation, as illustrated in Fig.~\ref{MAGIC_PASS framework}. The continuous tokens are decoded into multiple candidate configurations $\{\boldsymbol\Phi_{\rm t},\boldsymbol\phi_{\rm r},\mathbf W\}$ by a pretrained LLM. Moreover, the candidate configurations are parameterized to satisfy the PA placement and power budget. The candidate generation is trained offline. In the following subsections, we describe the candidate generation and offline training in detail.

\subsection{Candidate Generation}
The LLM output $\mathbf Z_{\mathcal E}^{\rm out}$ cannot be used directly as the optimized variables $(\boldsymbol\Phi_{\rm t},\boldsymbol\phi_{\rm r},\mathbf W)$ of \eqref{problem:P1}. Thus, we utilize a structured decoder comprising cross-attention decoders \cite{Vaswani2017Attention} to generate $N_{\rm cand}$ candidate configurations $\{\boldsymbol\Phi_{\rm t},\boldsymbol\phi_{\rm r},\mathbf W\}$ from the LLM output $\mathbf Z_{\mathcal E}^{\rm out}$. The decoder parameters are shared across candidates, but the candidate $i$ has a trainable vector $\mathbf q_i^{\rm cand}\in\mathbb R^{1\times D_{\rm M}}$ that is added to its decoder queries. Therefore, the candidates receive different query inputs even when $\mathbf Z_{\mathcal E}^{\rm out}$ is identical. Through three decoding functions $\operatorname{Dec}_{\rm t}(\cdot)$, $\operatorname{Dec}_{\rm r}(\cdot)$, and $\operatorname{Dec}_{\rm W}(\cdot)$ for transmit PA positions, receive PA position, and digital beamforming, the corresponding candidate features are expressed as follows:
\begin{subequations}
    \label{eq:structured_candidate_features}
    \begin{align}
        \mathbf F_{\rm t}^{(i)} & =\operatorname{Dec}_{\rm t}\!\left(\mathbf Q_{\rm t} + \mathbf 1_{(M_{\rm t}+1)N_{\rm t}}\mathbf q_i^{\rm cand}, \mathbf Z_{\mathcal E}^{\rm out}\right),\\
        \mathbf F_{\rm r}^{(i)} & = \operatorname{Dec}_{\rm r}\!\left(\mathbf Q_{\rm r}+\mathbf q_i^{\rm cand}, \mathbf Z_{\mathcal E}^{\rm out}\right),\\
        \mathbf F_{\rm W}^{(i)} & =\operatorname{Dec}_{\rm W}\!\left(\mathbf Q_{\rm W}+\mathbf 1_{K_{\rm c}}\mathbf q_i^{\rm cand},\mathbf Z_{\mathcal E}^{\rm out}\right),
    \end{align}
\end{subequations}
where $\mathbf 1_J$ denotes the length-$J$ all-one column vector, and the same candidate-identity vector $\mathbf q_i^{\rm cand}$ is added to the transmit-position, receive-position, and beamforming query banks. The structured decoder employs the trainable query banks $\mathbf Q_{\rm t}\in \mathbb R^{(M_{\rm t}+1)N_{\rm t}\times D_{\rm M}}$, $\mathbf Q_{\rm r}\in\mathbb R^{1\times D_{\rm M}}$, and $\mathbf Q_{\rm W}\in\mathbb R^{K_{\rm c}\times D_{\rm M}}$ for transmit PA positioning, receive PA positioning, and digital beamforming, respectively. The three outputs have the same row dimensions as their respective query banks. Because only one receive PA is used, both $\mathbf Q_{\rm r}$ and $\mathbf F_{\rm r}^{(i)}$ contain one row.

For the $n$-th transmit waveguide, a shared output head maps the corresponding rows of $\mathbf F_{\rm t}^{(i)}$ to $(M_{\rm t}+1)$ real-valued logits $\{a_{{\rm t},j,n}^{(i)}\}_{j=0}^{M_{\rm t}}$. Define $\pi_{{\rm t},j,n}^{(i)}=\frac{e^{(a_{{\rm t},j,n}^{(i)})}}{\sum_{r=0}^{M_{\rm t}}e^{(a_{{\rm t},r,n}^{(i)})}}$, where $j \in \{0,\ldots,M_{\rm t}\}$.
The resulting transmit-PA coordinate along the $x$-axis is given by
\begin{equation}
    \label{eq:transmit_pa_parameterization}
        x_{{\rm t},m,n}^{(i)} \!= \!x_{{\rm t},0,n} \!+\!(m\!-\!1)\Delta\!+\!\left[D_{\rm t}\!-\!(M_{\rm t}\!-\!1)\Delta\right]\sum_{j=0}^{m\!-\!1}\pi_{{\rm t},j,n}^{(i)},
\end{equation}
where $m\in [M_{\rm t}]$ and $n\in [N_{\rm t}]$. The $M_{\rm t}+1$ coefficients represent the nonnegative slack before, between, and after the $M_{\rm t}$ PAs. Provided that $D_{\rm t}\ge(M_{\rm t}-1)\Delta$, \eqref{eq:transmit_pa_parameterization} satisfies \eqref{transmit_position_constraint} and \eqref{transmit_spacing_constraint}. Specifically, for $m \ge 2$, $x_{{\rm t},m,n}^{(i)}-x_{{\rm t},(m-1),n}^{(i)}=\Delta+\left[D_{\rm t}-(M_{\rm t}-1)\Delta\right]\pi_{{\rm t},(m-1),n}^{(i)}\ge\Delta$. The terminal slack $\pi_{{\rm t},M_{\rm t},n}^{(i)}$ also ensures that the last PA does not exceed $x_{{\rm t},0,n}+D_{\rm t}$. Thus, no sorting or post-generation position projection is required.

The single receive-position feature produces the logit $a_{\rm r}^{(i)} = \mathbf F_{\rm r}^{(i)} \boldsymbol\theta_{\rm r}^{\rm pos} + b_{\rm r}^{\rm pos}\in\mathbb R$, where $\boldsymbol\theta_{\rm r}^{\rm pos}\in\mathbb R^{D_{\rm M}}$ and $b_{\rm r}^{\rm pos}\in\mathbb R$ are trainable. Its physical coordinate is given by
\begin{equation}
    x_{\rm r}^{(i)} = x_{{\rm r},0} + D_{\rm r}\operatorname{sigmoid}(a_{\rm r}^{(i)}),
    \label{eq:receive_pa_parameterization}
\end{equation}
where $\operatorname{sigmoid}(a)=1/(1+e^{-a})$ maps an unconstrained logit to $(0,1)$. Hence, \eqref{eq:receive_pa_parameterization} satisfies \eqref{receive_position_constraint} for every finite $a_{\rm r}^{(i)}$.
Using the fixed transverse coordinates, the three-dimensional PA positions are $\boldsymbol\phi_{{\rm t},m,n}^{(i)} = [x_{{\rm t},m,n}^{(i)},y_{{\rm t},n},z_{{\rm t},n}]^{\mathsf T}$, $\boldsymbol\phi_{\rm r}^{(i)} = [x_{\rm r}^{(i)},y_{\rm r},z_{\rm r}]^{\mathsf T}$, and $\boldsymbol\Phi_{\rm t}^{(i)} =[\boldsymbol\phi_{{\rm t},1,1}^{(i)},\ldots, \boldsymbol\phi_{{\rm t},M_{\rm t},N_{\rm t}}^{(i)}]$.

Then, the real-valued output head generates the real and imaginary parts of the unconstrained beamformer for each user as follows:
\begin{equation}
    \begin{bmatrix}
        \Re\{\widetilde{\mathbf w}_k^{(i)}\}\\
        \Im\{\widetilde{\mathbf w}_k^{(i)}\}
    \end{bmatrix}
    = \operatorname{Head}_{\rm W}\!([\mathbf F_{\rm W}^{(i)}]_{k,:}) \in\mathbb R^{2N_{\rm t}},\ k\in\mathcal K_{\rm c}.
\label{eq:beamforming_output_head}
\end{equation}

These vectors are collected as $\widetilde{\mathbf W}^{(i)}=[\widetilde{\mathbf w}_1^{(i)},\ldots,\widetilde{\mathbf w}_{K_{\rm c}}^{(i)}]$. Projection onto the feasible Frobenius-norm ball is expressed as follows:
\begin{equation}
    \mathbf W^{(i)} = \frac{\sqrt{P_{\max}}}{\max\{\sqrt{P_{\max}},\|\widetilde{\mathbf W}^{(i)}\|_{\rm F}\}}\widetilde{\mathbf W}^{(i)}.
\label{eq:beamforming_power_projection}
\end{equation}

This mapping leaves an already feasible matrix unchanged and otherwise scales it to satisfy \eqref{power_constraint} with equality.
The resulting LLM-generated candidate set is expressed as follows:
\begin{equation}
    \mathcal A_{\rm cand} = \{(\boldsymbol\Phi_{\rm t}^{(i)},\boldsymbol\phi_{\rm r}^{(i)},\mathbf W^{(i)})\}_{i=1}^{N_{\rm cand}},
    \label{eq:llm_candidate_set}
\end{equation}
where every element of $\mathcal A_{\rm cand}$ satisfies the transmit PA placement, receive PA placement, and power budget constraints specified in \eqref{transmit_position_constraint}-\eqref{power_constraint} by construction. It need not satisfy the QoS constraint \eqref{min_rate_constraint}, because the user rates depend jointly on the blockage-aware channels, PA positions, and multiuser interference and therefore cannot be guaranteed by independent output transformations. Consequently, $\mathcal A_{\rm cand}$ supplies feasible candidates for the subsequent refinement.

For the candidate $i$, the conditional diffusion model of the LLM explores a local region around each candidate, and thus, the candidate latent is expressed as follows:
\begin{equation}
    \mathbf v_{\rm anc}^{(i)} = [\operatorname{vec}^{\mathsf T}\!(\!\mathbf A_{\rm t}^{(i)}), a_{\rm r}^{(i)}, \operatorname{vec}^{\mathsf T}\!(\Re\{\widetilde{\mathbf W}^{(i)}\!\}), \operatorname{vec}^{\mathsf T}\!(\Im\{\widetilde{\mathbf W}^{(i)}\})]^{\mathsf T},
    \label{eq:configuration_anchor_latent}
\end{equation}
where $\mathbf v_{\rm anc}^{(i)}\in \mathbb{R}^{D_{\rm cfg}}$ and $D_{\rm cfg}=(M_{\rm t}+1)N_{\rm t}+1+2N_{\rm t}K_{\rm c}$ accounts for the transmit-position logits, one receive-position logit, and the real and imaginary beamforming entries. The mapping $\mathcal T_{\rm feas}(\mathbf v) = \big(\boldsymbol\Phi_{\rm t}(\mathbf v), \boldsymbol\phi_{\rm r}(\mathbf v), \mathbf W(\mathbf v)\big)$ follows \eqref{eq:transmit_pa_parameterization}, \eqref{eq:receive_pa_parameterization}, and \eqref{eq:beamforming_power_projection} and yields a hardware-feasible PASS configuration for every latent $\mathbf v$.
For a configuration latent $\mathbf v$, define the aggregate communication-QoS violation as follows:
\begin{equation}
    \mathcal V_{\rm QoS}(\mathbf v)\triangleq
    \sum_{k=1}^{K_{\rm c}}\left[R_k^{\min}-R_k\!\left(\mathbf U,\mathbf B,\boldsymbol\Phi_{\rm t}(\mathbf v),\mathbf W(\mathbf v)\right)\right]_+^2,
    \label{eq:aggregate_qos_violation}
\end{equation}
and the corresponding physics-guidance loss as
\begin{equation}
\begin{aligned}
    \mathcal L_{\rm phy}(\mathbf v)
    =&-R_{\rm r}\!\left(\boldsymbol\Phi_{\rm t}(\mathbf v),\mathbf u_{\rm tar},
    \boldsymbol\phi_{\rm r}(\mathbf v),\mathbf B,\mathbf W(\mathbf v)\right)\\
    &+\lambda_{\rm QoS}\mathcal V_{\rm QoS}(\mathbf v),
\end{aligned}
    \label{eq:physics_guidance_loss}
\end{equation}
where $\lambda_{\rm QoS}>0$ balances the sensing-rate objective and the QoS penalty. Its value is selected on a validation set and remains fixed online.

\subsection{Offline Training of MAGIC-PASS}

For each training environment $\mathcal E^{(q)}$, a multi-start AO/successive convex approximation (SCA) solver \cite{Scutari2017SCA,Zhang2025PASSISAC} is applied to \eqref{problem:P1}. The feasible solution with the highest sensing rate is retained as $\mathcal A^{\star(q)}=(\boldsymbol\Phi_{\rm t}^{\star(q)},\boldsymbol\phi_{\rm r}^{\star(q)},\mathbf W^{\star(q)})$. This numerical reference is not assumed to be globally optimal, and the solver is used only during offline training.
The numerical solver returns PA positions and a beamforming matrix, whereas the trainable modules use the latent representation in \eqref{eq:configuration_anchor_latent}. Therefore, each $\mathcal A^{\star(q)}$ is converted into a reference latent $\mathbf v^{\star(q)}$ using a fixed inverse convention. The transmit-PA positions are converted into the logits used in \eqref{eq:transmit_pa_parameterization}, the receive-PA position is converted using the inverse of \eqref{eq:receive_pa_parameterization}, and the feasible $\mathbf W^{\star(q)}$ is used directly before power projection. A small guard keeps the position logits finite at the PA-placement boundaries. The resulting training set is $\mathcal D_{\rm tr} =\{(\mathcal E^{(q)},\mathbf v^{\star(q)})\}_{q=1}^{N_{\rm tr}}$. 
The offline training of MAGIC-PASS is divided into three stages: i) multimodal tokenization modules and candidate generator training, ii) conditional denoiser training, and iii) joint optimization of all modules.

In the first stage, for each $\mathcal E^{(q)}$, the generator produces $N_{\rm cand}$ candidate latents $\{\mathbf v_{\rm anc}^{(q,i)}\}_{i=1}^{N_{\rm cand}}$. The training data, however, contain only one solution $\mathbf v^{\star(q)}$ for that environment. Thus, we measure the distance between each candidate and the training solution as $d_{q,i}\triangleq \|\boldsymbol\Gamma_{\rm cfg} (\mathbf v_{\rm anc}^{(q,i)}-\mathbf v^{\star(q)})\|_2^2$ and minimize the best-of-$N_{\rm cand}$ loss $\mathcal L_{\rm cand}=\frac{1}{N_{\rm tr}}\sum_{q=1}^{N_{\rm tr}}\min_{i\in[N_{\rm cand}]}d_{q,i}$,
where $\boldsymbol\Gamma_{\rm cfg}$ is a fixed diagonal matrix that places the transmit-PA position, receive-PA position, and beamforming variables on comparable scales. Let $i_q^\star$ denote a candidate attaining the minimum. Only this candidate is matched to the feasible solutions by $\mathcal L_{\rm cand}$, the remaining candidates are still trained by the following two losses:

1) The diversity loss penalizes candidate pairs whose normalized latent distance is below $\delta_{\rm div}>0$. With $\Delta\mathbf v_{q,i,j} \triangleq\mathbf v_{\rm anc}^{(q,i)}-\mathbf v_{\rm anc}^{(q,j)}$, the diversity regularizing loss penalizes a pair only when its normalized latent distance is smaller than $\delta_{\rm div}$, defined by
\begin{equation}
\begin{aligned}
    \mathcal L_{\rm div} \!=\!\frac{2}{N_{\rm tr}N_{\rm cand}\!(\!N_{\rm cand}\!-\!1\!)}\!\sum_{q=1}^{N_{\rm tr}}\!\sum_{1\!\le \!i<\!j\!\le\! N_{\rm cand}\!}\![\delta_{\rm div}\!-\! \|\boldsymbol\Gamma_{\rm cfg}\Delta\mathbf v_{q,i,j}\|_2]_+^2,
\end{aligned}
\label{eq:candidate_diversity_loss}
\end{equation}
where $\mathcal L_{\rm div}=0$ as $N_{\rm cand}=1$.

2) For environment $\mathcal E^{(q)}$, let $\mathcal L_{\rm phy}^{(q)}$ denote the configuration loss to evaluate the performance of every candidate using that environment's user positions, target position, blockage state, and rate requirements. Unlike the candidate-reference loss, the candidate physics loss applies to all candidates, expressed as $\mathcal L_{\rm phy}^{\rm anc}=\frac{1}{N_{\rm tr}N_{\rm cand}}\sum_{q=1}^{N_{\rm tr}}\sum_{i=1}^{N_{\rm cand}}\mathcal L_{\rm phy}^{(q)}(\mathbf v_{\rm anc}^{(q,i)})$.

In the second stage, after training the candidate generator, we pair each feasible solution $\mathbf v^{\star(q)}$ with its closest generated candidate $\mathbf v_{\rm anc}^{(q,i_q^\star)}$ to train the conditional denoiser. A diffusion index $\ell$ is sampled uniformly from $[N_{\rm diff}]$, and Gaussian noise $\boldsymbol\omega\sim\mathcal N(\mathbf0,\mathbf I_{D_{\rm cfg}})$ is added to the feasible solution according to the following:
\begin{equation}
    \label{eq:forward_configuration_diffusion}
    \mathbf v_\ell^{\star(q)}=\sqrt{\bar\alpha_\ell}\mathbf v^{\star(q)}+\sqrt{1-\bar\alpha_\ell}\boldsymbol\omega.
\end{equation}

This forward process adds Gaussian noise to the feasible solution $\mathbf v^{\star(q)}$. Then, the denoiser is trained to estimate the added noise from the perturbed reference, diffusion index, environment representation, and paired candidate \cite{Ho2020DDPM,Song2021Score}, and the prediction error is measured by
\begin{equation}
    \begin{aligned}
    \mathcal L_{\rm diff}=\mathbb E[\|\boldsymbol\omega -\operatorname{Dn}_{\boldsymbol\Psi}(\mathbf v_\ell^{\star(q)},\ell,\mathbf Z_{\mathcal E^{(q)}}^{\rm out}, \mathbf v_{\rm anc}^{(q,i_q^\star)})\|_2^2].
    \end{aligned}
    \label{eq:conditional_diffusion_loss}
\end{equation}

In the third stage, all modules are jointly optimized by the complete offline training loss, expressed as follows:
\begin{equation}
    \mathcal L_{\rm off}=\mathcal L_{\rm cand}+\lambda_{\rm div}\mathcal L_{\rm div}+\lambda_{\rm phy}^{\rm tr}\mathcal L_{\rm phy}^{\rm anc}+\lambda_{\rm diff}\mathcal L_{\rm diff},
    \label{eq:overall_offline_training_loss}
\end{equation}
where $\lambda_{\rm div}$, $\lambda_{\rm phy}^{\rm tr}$, and $\lambda_{\rm diff}$ are nonnegative validation-selected weights.

\section{LLM-Assisted PASS Configuration Refinement}
\label{subsec:diffusion_refinement}

This section refines the generated candidate configurations online using the trained conditional denoiser and physical guidance. The resulting configurations are then evaluated to select the final configuration under the communication-QoS requirements. Additionally, the computational complexity of the online optimization is analyzed.

\subsection{Refinement and Selection}
For the environment $\mathcal E$, the candidate configurations generated in Section~\ref{subsec:candidate_configuration_generation} are refined to select the configuration with the highest sensing-rate using the conditional denoiser with trained parameters and the physical loss in \eqref{eq:physics_guidance_loss}. All trained model parameters remain fixed during this process. After the prescribed refinement steps, the refined configurations are evaluated for their sensing rates and communication-QoS requirements. Among feasible configurations, the one with the highest sensing rate is selected.


During the online refinement, each candidate is initialized by partially perturbing its latent rather than sampling from pure Gaussian noise, which is expressed as follows:
\begin{equation}
    \mathbf v_{\ell_0}^{(i)} = \sqrt{\bar\alpha_{\ell_0}}\mathbf v_{\rm anc}^{(i)}+\sqrt{1\!-\!\bar\alpha_{\ell_0}}\boldsymbol\omega^{(i)},\ \boldsymbol\omega^{(i)}\sim \mathcal N(\mathbf0,\mathbf I_{D_{\rm cfg}}),
    \label{eq:anchor_partial_perturbation}
\end{equation}
where $i\in[N_{\rm cand}]$, the prescribed starting index $\ell_0\in[N_{\rm diff}]$ controls the exploration radius, $N_{\rm diff}$ denotes the number of diffusion steps, $\nu_\ell\in(0,1)$ is the variance at step $\ell$, $\alpha_\ell=1-\nu_\ell$, and $\bar\alpha_\ell=\prod_{r=1}^{\ell}\alpha_r$. A smaller $\ell_0$ retains more latent information, whereas a larger $\ell_0$ permits broader exploration. Its value is selected on the validation set and fixed during online optimization. The conditional denoiser $\operatorname{Dn}_{\boldsymbol\Psi}(\cdot)$ estimates the noise in a perturbed configuration latent using the diffusion index, the environment representation $\mathbf Z_{\mathcal E}^{\rm out}$, and the candidate latent $\mathbf v_{\rm anc}$. For $\ell=\ell_0,\ldots,1$, the learned reverse update is computed by the conditional denoiser, expressed as follows:
\begin{equation}
        \overline{\mathbf v}_{\ell\!-\!1}^{(i)}\!=\!\frac{1}{\sqrt{\alpha_\ell}}[\mathbf v_\ell^{\!(i)\!}\!-\!\frac{\nu_\ell}{\sqrt{1\!-\!\bar\alpha_\ell}}\operatorname{Dn}_{\boldsymbol\Psi}\!(\!\mathbf v_\ell^{(i)},\ell,\mathbf Z_{\mathcal E}^{\rm out},\mathbf v_{\rm anc}^{(i)}\! )\!]\!+\!\sqrt{\widetilde\nu_\ell}\mathbf z_\ell^{\!(i)\!},
    \label{eq:conditional_reverse_diffusion}
\end{equation}
where $\widetilde\nu_\ell \triangleq \frac{\nu_\ell(1-\bar\alpha_{\ell-1})}{1-\bar\alpha_\ell}$, $\bar\alpha_0=1$, $\mathbf z_\ell^{(i)} \sim\mathcal N(\mathbf0,\mathbf I_{D_{\rm cfg}})$ for $\ell>1$, and $\mathbf z_1^{(i)}=\mathbf0$. This reverse update follows the learned conditional distribution, but it does not explicitly guarantee communication QoS. To align the refined configurations with the communication-QoS requirements, the latent is then adjusted under \eqref{eq:aggregate_qos_violation} and \eqref{eq:physics_guidance_loss}, given by
\begin{equation}
    \mathbf v_{\ell-1}^{(i)}=\overline{\mathbf v}_{\ell-1}^{(i)}-\mu_\ell\nabla_{\overline{\mathbf v}_{\ell-1}^{(i)}}\mathcal L_{\rm phy}\!\left(\overline{\mathbf v}_{\ell-1}^{(i)}\right),
\label{eq:physics_guided_reverse_update}
\end{equation}
where $\mu_\ell>0$ is the prescribed guidance step size \cite{Dhariwal2021Diffusion}. To evaluate the gradient $\nabla\mathcal L_{\rm phy}$, $\mathcal T_{\rm feas}(\cdot)$ maps the latent to PA positions and a beamforming matrix, from which the rates in \eqref{eq:physics_guidance_loss} are computed.
With a prescribed tolerance $\epsilon_{\rm QoS}>0$, define the set of QoS-feasible candidate indices as follows:
\begin{equation}
    \mathcal I_{\rm QoS} \triangleq \{ i\in[N_{\rm cand}]\ |\ \mathcal V_{\rm QoS}(\mathbf v_{0}^{(i)})\le\epsilon_{\rm QoS} \},
    \label{eq:qos_feasible_candidate_set}
\end{equation}
where this set allows a rate shortfall of at most $\sqrt{\epsilon_{\rm QoS}}$ for each user. If $\mathcal I_{\rm QoS}\neq\varnothing$, we select the candidate with the highest sensing rate among those satisfying the QoS requirement:
\begin{equation}
    i^\star\in\underset{i\in\mathcal I_{\rm QoS}}{\arg\max}\;R_{\rm r}(\widehat{\boldsymbol\Phi}_{\rm t}^{(i)},\mathbf u_{\rm tar},\widehat{\boldsymbol\phi}_{\rm r}^{(i)},\mathbf B,\widehat{\mathbf W}^{(i)}).
    \label{eq:online_final_candidate_selection}
\end{equation}

Then, the $i$-th refined configuration is given by
\begin{equation}
    \widehat{\mathcal A}^{(i)}\triangleq (\widehat{\boldsymbol\Phi}_{\rm t}^{(i)},\widehat{\boldsymbol\phi}_{\rm r}^{(i)},\widehat{\mathbf W}^{(i)})=\mathcal T_{\rm feas}(\mathbf v_0^{(i)}),
    \label{eq:refined_physical_candidate}
\end{equation}
where the candidate $\widehat{\mathcal A}_{\rm cand}\triangleq\{\widehat{\mathcal A}^{(i)}\}_{i=1}^{N_{\rm cand}}$ is refined to satisfy the explicit QoS and to finish the final selection procedure.

If no refined candidate satisfies the QoS requirement, we select the one with the smallest rate violation $i_0\in\underset{i\in[N_{\rm cand}]}{\arg\min}\;\mathcal V_{\rm QoS}(\mathbf v_0^{(i)})$.
Starting from $\mathbf v_{\rm corr}^{(0)}=\mathbf v_0^{(i_0)}$, we update its latent to reduce the rate violation, given by
\begin{equation}
        \mathbf v_{\rm corr}^{(j+1)}=\mathbf v_{\rm corr}^{(j)}-\gamma_j\nabla_{\mathbf v_{\rm corr}^{(j)}}\mathcal V_{\rm QoS}\!(\mathbf v_{\rm corr}^{(j)}),\ \gamma_j>0.
    \label{eq:feasibility_correction_update}
\end{equation}

The correction stops when the QoS requirement is satisfied or the budget $N_{\rm corr}$ is exhausted.
We retain the iteration with the smallest rate violation and use the sensing rate to break ties. The online procedure is summarized in Algorithm~\ref{alg:online_magic_pass}.

\begin{algorithm}[t]
\caption{Online LLM-Assisted Generative Optimization}
\label{alg:online_magic_pass}
\begin{algorithmic}[1]
\REQUIRE Fixed trained model; current environment $\mathcal E$; system
parameters; $N_{\rm cand}$, $\ell_0$, $N_{\rm corr}$;
$\{\mu_\ell\}_{\ell=1}^{\ell_0}$,
$\{\gamma_j\}_{j=0}^{N_{\rm corr}-1}$; $\epsilon_{\rm QoS}$.
\ENSURE PASS configuration and feasibility status.

\STATE Construct $\mathbf{Z}_{\mathcal{E}}^{\rm in}$, evaluate the LLM once,
and cache $\mathbf{Z}_{\mathcal{E}}^{\rm out}$.
\STATE Generate $\{\mathbf v_{\rm anc}^{(i)}\}_{i=1}^{N_{\rm cand}}$ and
$\mathcal A_{\rm cand}$.

\FOR{$i=1,\ldots,N_{\rm cand}$ \textbf{in parallel}}
    \STATE Initialize $\mathbf v_{\ell_0}^{(i)}$ using
    \eqref{eq:anchor_partial_perturbation}.
    \FOR{$\ell=\ell_0,\ldots,1$}
        \STATE Apply the updates in
        \eqref{eq:conditional_reverse_diffusion} and
        \eqref{eq:physics_guided_reverse_update}.
    \ENDFOR
    \STATE Obtain $\widehat{\mathcal A}^{(i)} =\mathcal T_{\rm feas}(\mathbf v_0^{(i)})$.
\ENDFOR

\STATE Construct $\mathcal I_{\rm QoS}$ using \eqref{eq:qos_feasible_candidate_set}.

\IF{$\mathcal{I}_{\rm QoS}\neq\varnothing$}
    \STATE Select $i^\star$ using
    \eqref{eq:online_final_candidate_selection}.
    \STATE \textbf{return} $\widehat{\mathcal A}^{(i^\star)}$.
\ELSE
    \STATE Select
    $i_0\in\arg\min_{i\in[N_{\rm cand}]}\mathcal V_{\rm QoS}(\mathbf v_0^{(i)})$.
    \STATE Set $\mathbf v_{\rm corr}^{(0)}=\mathbf v_{\rm best}=\mathbf v_0^{(i_0)}$.

    \FOR{$j=0,\ldots,N_{\rm corr}-1$}
        \STATE Compute $\mathbf v_{\rm corr}^{(j+1)}$ using \eqref{eq:feasibility_correction_update}.
        \STATE Update $\mathbf v_{\rm best}$ if the new iterate has a smaller violation; break ties using the sensing rate.
        \IF{$\mathcal V_{\rm QoS}(\mathbf v_{\rm corr}^{(j+1)})
        \le\epsilon_{\rm QoS}$}
            \STATE \textbf{return}
            $\mathcal T_{\rm feas}(\mathbf v_{\rm corr}^{(j+1)})$.
        \ENDIF
    \ENDFOR
    \STATE \textbf{return} $\mathcal T_{\rm feas}(\mathbf v_{\rm best})$.
\ENDIF
\end{algorithmic}
\end{algorithm}

\subsection{Computational Complexity of Online Implementation}
\label{subsec:complexity_online_implementation}

The online computation comprises multimodal feature processing, one frozen-LLM forward pass, structured candidate generation, diffusion refinement, and optional feasibility correction. Let $\Xi_{\rm enc}$ denote the total cost of the modality-specific encoders and projection modules. For an LLM with $N_{\rm LLM}$ transformer layers, $N_{\mathcal E}$ input tokens, and hidden dimension $D_{\rm M}$, its forward-pass complexity is $\Xi_{\rm LLM}=\mathcal O\!\left(N_{\rm LLM}(N_{\mathcal E}^{2}D_{\rm M}+N_{\mathcal E}D_{\rm M}^{2})\right)$. This cost is incurred once per environment because $\mathbf Z_{\mathcal E}^{\rm out}$ is shared by all candidate branches.

The query numbers of the three structured decoders are $N_{{\rm q},{\rm t}}=(M_{\rm t}+1)N_{\rm t}$, $N_{{\rm q},{\rm r}}=1$, $N_{{\rm q},{\rm W}}=K_{\rm c}$. Since the transmit-position, receive-position, and beamforming decoders in \eqref{eq:structured_candidate_features} are independent, their self-attention costs must be evaluated separately. For $N_{\rm dec}$ decoder layers, the cost of generating one candidate is $\Xi_{\rm dec}\!=\!\mathcal O(N_{\rm dec}\!\sum_{\!p\in\{{\rm t},{\rm r},{\rm W}\}}\!(\!N_{{\rm q},p}^{2}D_{\rm M}\!+\!N_{{\rm q},p}N_{\mathcal E}D_{\rm M}\!+\!N_{{\rm q},p}D_{\rm M}^{2})\!)$,
where the three terms represent query self-attention, cross-attention with the
cached LLM representation, and feedforward processing, respectively.

Let $\Xi_{\rm den}$ denote the cost of one conditional-denoiser evaluation and $\Xi_{\rm phy}$ the cost of evaluating and differentiating $\mathcal L_{\rm phy}$. Refining all candidates requires $N_{\rm cand}\ell_0(\Xi_{\rm den}+\Xi_{\rm phy})$ operations. In addition, let $\Xi_{\rm QoS}$ denote the cost of evaluating and differentiating $\mathcal V_{\rm QoS}$. Since the optional correction is applied to one selected latent, it requires at most $N_{\rm corr}\Xi_{\rm QoS}$ operations. The resulting worst-case online workload is given by
\begin{equation}
    \begin{aligned}
        \Xi_{\rm on}={}&\Xi_{\rm enc}+\Xi_{\rm LLM}+N_{\rm cand}\Xi_{\rm dec}\\
        & + N_{\rm cand}\ell_0(\Xi_{\rm den}+\Xi_{\rm phy})+N_{\rm corr}\Xi_{\rm QoS},
    \end{aligned}
    \label{eq:overall_online_complexity}
\end{equation}
where the last term $N_{\rm corr}\Xi_{\rm QoS}$ is incurred only when none of the refined candidates satisfies \eqref{eq:qos_feasible_candidate_set}. The lower-order candidate-screening and selection operations are omitted from \eqref{eq:overall_online_complexity}, and \eqref{eq:overall_online_complexity} describes total arithmetic workload. Although this workload scales linearly with $N_{\rm cand}$, the candidate branches share the cached LLM representation and can be processed as one batch. Parallel execution can therefore reduce wall-clock latency, but does not reduce the total number of arithmetic operations. Excluding fixed model parameters and temporary decoder or automatic-differentiation activations, the persistent cached-state memory is $\mathcal O(N_{\mathcal E}D_{\rm M}+N_{\rm cand}D_{\rm cfg})$.

\section{Numerical Results}
\label{simulation_results}

This section evaluates the convergence, sensing--communication performance, and component effectiveness of the proposed MAGIC-PASS. Unless otherwise stated, all schemes are evaluated over the same test scenarios and under their respective common physical constraints. The principal simulation and learning parameters are listed in Table~\ref{tab:simulation_parameters}.

\begin{table}[t]
\centering
\caption{Principal simulation and learning parameters.}
\label{tab:simulation_parameters}
\footnotesize
\setlength{\tabcolsep}{3.5pt}
\renewcommand{\arraystretch}{0.96}
\begin{tabular}{@{}p{0.50\columnwidth}p{0.36\columnwidth}@{}}
\hline
Parameter(s) & Value(s) \\
\hline
$f_{\rm c}$; region; vertical separation
    & $28$~GHz; $20\times20$~m$^2$; $3$~m \\
$N_{\rm t}$; $M_{\rm t}$; $K_{\rm c}$
    & $4$; $4$; $4$ \\
$(D_{\rm t},D_{\rm r})$; $\Delta$
    & $(20,20)$~m; $\lambda/2$ \\
$n_{\rm eff}$; $\kappa$
    & $1.4$; $0.1$~dB/m \\
$P_{\max}$; $(\sigma_{\rm c}^2,\sigma_{\rm s}^2)$
    & $25$~dBm; $\!(-90,\!-120)\!$~dBm \\
$L$; $\alpha_{\rm s}$; $\epsilon_{\rm QoS}$
    & $64$; $1$; $10^{-6}$ \\
$(\eta_{\rm o},\eta_{\rm CSI})$; $R_k^{\min}$
    & $(5,5)$; $2$--$12$~bit/s/Hz \\
Default $P_{\max}$; swept $P_{\max}$
    & $25$~dBm; $10$--$40$~dBm \\
Default $R^{\min}$; swept $R^{\min}$
    & $12$~bit/s/Hz; $2$--$12$~bit/s/Hz \\
LLM backbone; $D_{\rm M}$
    & Qwen2.5-1.5B; $1536$ \\
$N_{\rm cand}$; $(N_{\rm diff},\ell_0,N_{\rm corr})$
    & $4$; $(50,10,20)$ \\
Batch size; optimizer; learning rate
    & $16$; AdamW \cite{Loshchilov2019AdamW}; $10^{-4}$ \\
Maximum epochs (three stages)
    & $80/80/40$ \\
Train/validation/test samples; seeds
    & $8000/1000/1000$; $3$ \\
\hline
\end{tabular}
\end{table}

\subsection{Simulation Setup}

We consider the downlink PASS-enabled ISAC system in Section~\ref{Section: System_Model}, operating at $f_{\rm c}=28$~GHz over a $20\times20$~m$^2$ service region. The BS employs $N_{\rm t}=4$ transmit waveguides, each activating $M_{\rm t}=4$ PAs, and one receive waveguide with one active PA. Four single-antenna users and one target are independently and uniformly placed in the service region, $3$~m below the waveguides. The BS serves all $K_{\rm c}=4$ users while reusing the downlink signal to estimate the reflection coefficient of the target \cite{Guo2025Learning}.

The representation backbone is the frozen Qwen2.5-1.5B model \cite{qwen2025qwen25}, whose hidden dimension is $D_{\rm M}=1536$. Images are resized to $224\times224$ pixels and processed by a frozen CLIP ViT-B/32 encoder \cite{radford2021learning}. These are the only frozen encoders. A trainable PointNet encoder with feature dimensions 64, 128, and 256 produces eight LiDAR tokens \cite{Qi2017PointNet}, while trainable two-layer multilayer perceptrons encode the location, CSI, blockage, and task inputs. The modality-projection modules, structured decoders, and conditional denoiser are also trained offline. Each of the three structured decoders has two cross-attention layers, eight heads, and a feedforward dimension of 2048. The conditional denoiser contains four residual blocks of width 1024. Under the stated system dimensions, $D_{\rm cfg}=(M_{\rm t}+1)N_{\rm t}+1+2N_{\rm t}K_{\rm c}=53$. The trainable task-specific modules are configured and trained for stated values of $N_{\rm t}$, $M_{\rm t}$, and $K_{\rm c}$. Changing these dimensions requires resizing and retraining the affected input and output modules, while the frozen LLM backbone can be retained. Here, adaptability refers to generating a configuration for each tested user--target geometry and blockage state under fixed $N_{\rm t}$, $M_{\rm t}$, and $K_{\rm c}$.

The diffusion process uses $N_{\rm diff}=50$ steps and a linear variance schedule from $10^{-4}$ to $2\times10^{-2}$. Online refinement starts from $\ell_0=10$ and permits at most $N_{\rm corr}=20$ feasibility-correction iterations. The QoS tolerance is $\epsilon_{\rm QoS}=10^{-6}$. The loss coefficients are $\lambda_{\rm div}=0.05$, $\lambda_{\rm phy}^{\rm tr}=0.1$, $\lambda_{\rm diff}=1$, and $\lambda_{\rm QoS}=10$, with $\delta_{\rm div}=7.25$. The guidance and correction step-size schedules, together with all loss hyperparameters, are selected using the validation set and then fixed for testing.

All modalities in one sample correspond to the same physical scenario. In particular, the user and target positions, visual and LiDAR observations, blockage descriptor, and historical CSI are aligned to a common environmental realization; the historical CSI is paired with the PA configurations used for its acquisition. The task token contains the user-rate requirements and power budget defined in Section~\ref{Section: System_Model}.

For each training scenario, five independently initialized AO/SCA runs are performed to generate the numerical reference. Each run terminates after 100 iterations or when the relative objective improvement is below $10^{-4}$ \cite{Scutari2017SCA,Zhang2025PASSISAC}. The feasible solution with the largest sensing rate is converted into the canonical reference latent $\mathbf v^\star$ using the procedure in Section~\ref{subsec:candidate_configuration_generation}. The reported dataset sizes count the retained scenarios for which at least one feasible numerical reference is obtained. The three offline training stages use at most 80, 80, and 40 epochs, respectively, with early stopping determined from the validation loss.

\subsection{Benchmark Schemes}

The proposed MAGIC-PASS is compared with the following three benchmarks.

\emph{1) Con-MIMO:} This benchmark employs a conventional fixed-array hybrid MIMO-ISAC architecture. Its transmit array contains $M_{\rm t}N_{\rm t}=16$ half-wavelength-spaced elements driven by $N_{\rm t}=4$ RF chains, while one fixed receive element is placed at the center of the receive-PA movement interval. The constant-modulus analog precoder and digital precoder are optimized by AO/SCA under the same total transmit-power and communication-QoS constraints, following \cite{wang2025hybrid}.

\emph{2) LLM-MIMO:} This scheme applies LLM-assisted beamforming to the same fixed MIMO hardware as Con-MIMO. It uses the same multimodal inputs, frozen backbone, scenario split, and candidate budget as MAGIC-PASS, but optimizes only the hybrid beamforming variables and uses candidate generation without diffusion refinement to optimize hybrid beamforming while keeping the antenna positions fixed. Its design follows the LLM-assisted beamforming principle in \cite{wang2026beamagent}.

\emph{3) Con-PASS:} This conventional optimization-based scheme uses the same PASS hardware as MAGIC-PASS and alternately updates $\mathbf W$, $\boldsymbol\Phi_{\rm t}$, and the single receive-PA position $\boldsymbol\phi_{\rm r}$. The resulting nonconvex subproblems are handled by SCA under the same power, QoS, movement-range, and transmit-PA spacing constraints, following \cite{Zhang2025PASSISAC}.


For the hardware-level comparison, all schemes employ 16 transmit radiating elements, four transmit RF chains, one sensing receive element, and the same total transmit-power budget. They are evaluated under identical user and target geometries, blockage realizations, noise powers, and QoS requirements. MAGIC-PASS and LLM-MIMO use the same training, validation, and test scenario partitions and the same online candidate budget, but optimize configurations for their respective PASS and fixed-array hardware. Con-PASS uses one initialization online; the multi-start AO/SCA procedure used to construct the training references for MAGIC-PASS is not included in the Con-PASS online benchmark. LLM-MIMO does not perform diffusion refinement, whereas MAGIC-PASS uses the prescribed reverse-diffusion steps and QoS correction. Therefore, schemes have different online workloads, and the sensing-rate comparison should not be interpreted as a matched-latency comparison.

\subsection{Convergence Behaviors}

\begin{figure}[t]
    \centering
    \includegraphics[width=0.8\columnwidth]{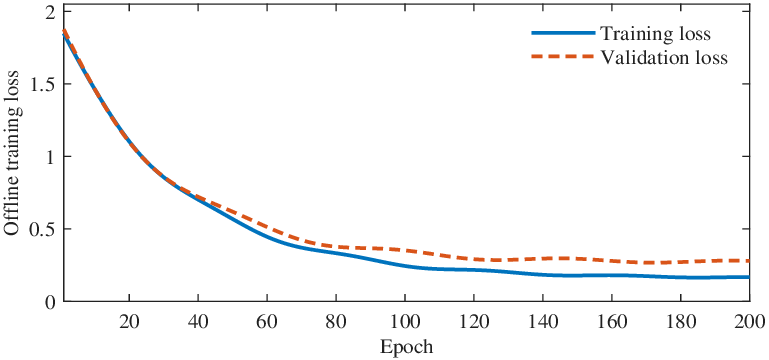}
    \caption{Stage-wise training and validation losses. The stage boundaries
    occur at epochs 80 and 160.}
    \label{offline_training}
\end{figure}

Fig.~\ref{offline_training} presents the stage-wise offline training behavior.
Epochs 1--80 train the modality-alignment and candidate-generation modules,
epochs 81--160 train the conditional denoiser with the candidate path fixed,
and epochs 161--200 perform joint optimization of all trainable non-backbone
modules. Because the active objective changes between stages, absolute loss
values should be compared within, rather than across, the three stages. In
each stage, the validation curve follows the training curve and levels off
before the corresponding epoch budget is exhausted. During the final joint
stage, the training loss approaches approximately $0.16$, while the
validation loss remains below $0.3$, without evident late-epoch divergence.

\begin{figure}[t]
    \centering
    \includegraphics[width=0.8\columnwidth]{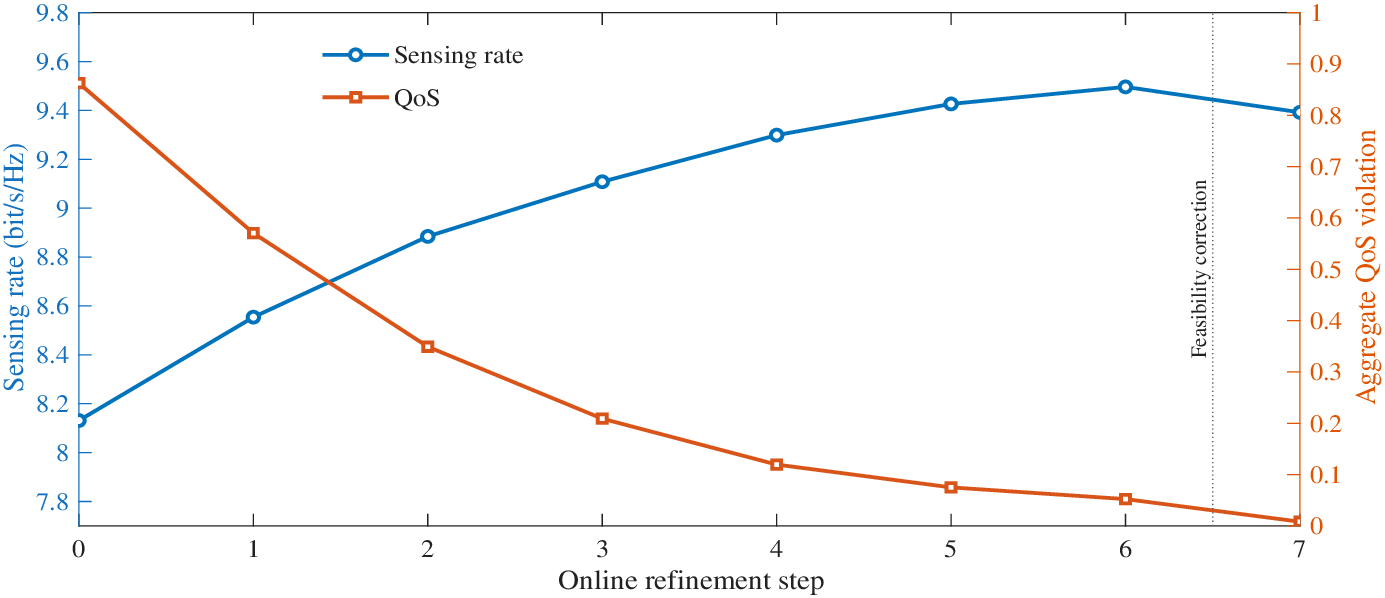}
    \caption{Average online refinement trajectory for $\ell_0=10$, showing
    selected reverse-diffusion checkpoints and the feasibility-correction
    output.}
    \label{online_refinement}
\end{figure}

Fig.~\ref{online_refinement} presents the average online trajectory for $\ell_0=10$. To avoid overcrowding, selected checkpoints from the ten reverse-diffusion steps are displayed together with the correction output. Before reverse refinement, the generated configurations attain approximately $8.1$~bit/s/Hz and an aggregate QoS violation of about $0.85$. At the last displayed reverse checkpoint, the sensing rate reaches approximately $9.5$~bit/s/Hz, while the violation decreases to about $0.05$. Thus, the conditional reverse updates and physical guidance jointly improve the sensing objective and recover most of the communication deficit. The subsequent QoS-oriented correction drives the residual violation below $\epsilon_{\rm QoS}$, with a small sensing-rate reduction to approximately $9.4$~bit/s/Hz. This reduction is expected because the fallback correction prioritizes feasibility rather than further sensing-rate maximization.

\subsection{Sensing and Communication Performances}

\begin{figure}[t]
    \centering
    \includegraphics[width=0.8\columnwidth]{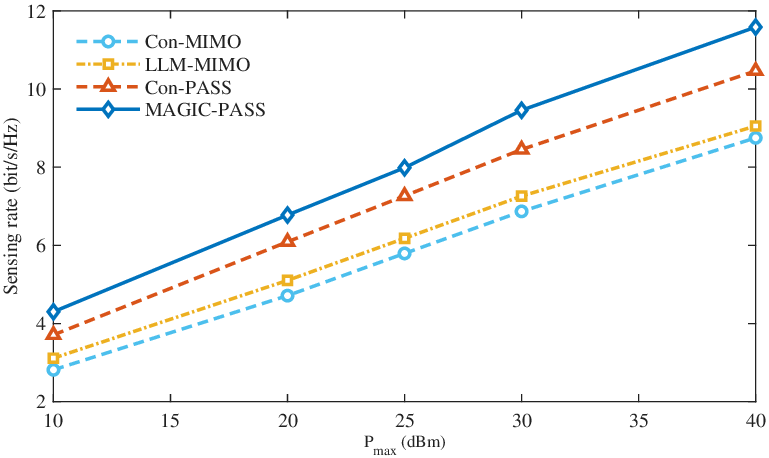}
    \caption{Sensing rate versus the power budget $P_{\max}$ under a
    fixed common user-rate requirement.}\label{sensing_power}
\end{figure}

Fig.~\ref{sensing_power} compares the sensing rate under different transmit-power budgets, with the common user-rate requirement held fixed for all schemes. The sensing rate increases monotonically with $P_{\max}$ because a larger power budget strengthens the received echo and provides greater flexibility in meeting the communication QoS constraints. The LLM-MIMO consistently outperforms the Con-MIMO, demonstrating the benefit of multimodal beamforming with fixed arrays. The Con-PASS performs better than both MIMO benchmarks by exploiting the additional spatial degrees of freedom provided by PA repositioning. MAGIC-PASS achieves the highest sensing rate over the entire power range, showing the further benefit of multimodal candidate generation and refinement. At $P_{\max}=40$~dBm, MAGIC-PASS achieves approximately $11.3$~bit/s/Hz, compared with $10.5$, $9.0$, and $8.8$~bit/s/Hz for the Con-PASS, the LLM-MIMO, and the Con-MIMO, respectively. These values correspond to gains of $7.6\%$, $25.6\%$, and $28.4\%$.

\begin{figure}[t]
    \centering
    \subfloat[]
    {\includegraphics[width=0.8\columnwidth]{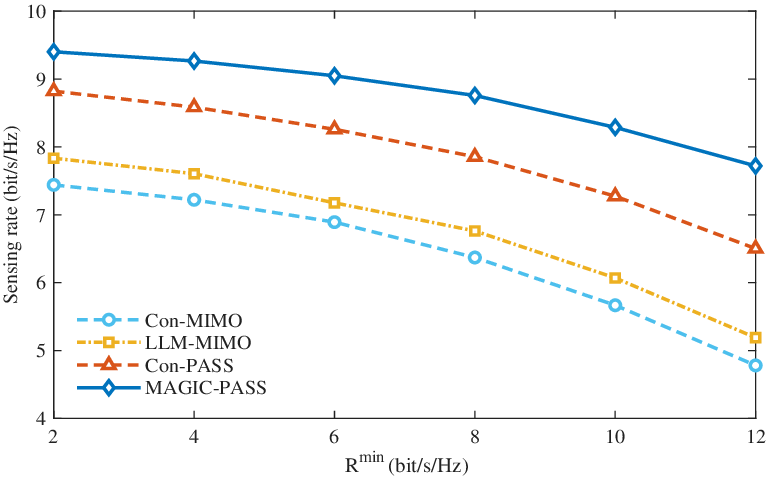}
    \label{sensing_tradeoff}}
    \hfill
    \subfloat[]
    {\includegraphics[width=0.8\columnwidth]{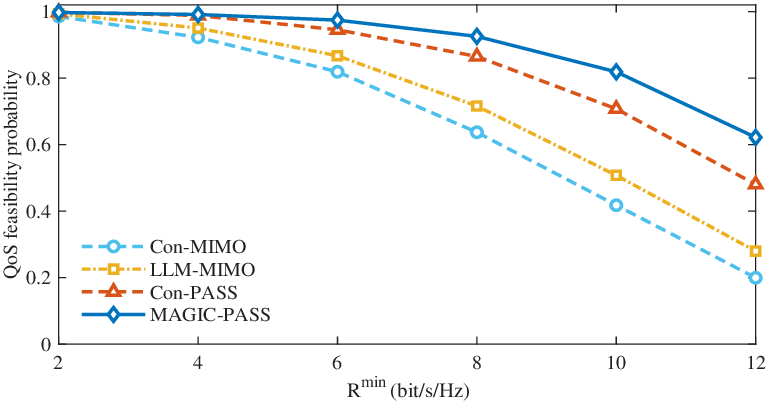}
    \label{qos_tradeoff}}
    \caption{Sensing-communication trade-off versus the minimum user-rate requirement: (a) sensing rate and (b) QoS-feasibility probability.}\label{tradeoff}
\end{figure}

Fig.~\ref{tradeoff} evaluates the sensing-communication tradeoff as the common user-rate requirement $R_k^{\min}=R^{\min}$ increases. For every method, the sensing rate is computed over all $N_{\rm test}$ test scenarios. Hence, the reported averages do not exclude QoS-infeasible outputs. Using the same tolerance $\epsilon_{\rm QoS}$ for all schemes, the QoS-feasibility probability is defined as $P_{\rm feas} \triangleq \frac{1}{N_{\rm test}} \sum_{q=1}^{N_{\rm test}} \mathbf 1\!\left\{\mathcal V_{\rm QoS}^{(q)}\leq\epsilon_{\rm QoS}\right\}$, where $\mathbf 1\{\cdot\}$ is the indicator function. Both the sensing rate and $P_{\rm feas}$ decrease with $R^{\min}$ because increasingly stringent communication requirements leave less flexibility for sensing and make joint feasibility more difficult.

MAGIC-PASS maintains the highest sensing rate and QoS-feasibility probability over the entire range. Across the considered user-rate requirements, MAGIC-PASS achieves higher average sensing rates and QoS-feasibility probabilities than the three benchmarks. Compared with the fixed-array MIMO schemes, PASS can adjust the transmit and receive PA positions to improve the user links and the target echo. MAGIC-PASS also outperforms Con-PASS, although both use the same PASS settings, which indicates that candidate generation followed by physical refinement yields more effective configurations than the conventional optimization method. Moreover, MAGIC-PASS utilizes the multimodal environmental information to improve the adaptability of PASS configurations to varied user/target locations and blockage conditions compared with LLM-MIMO.

\begin{figure}[t]
    \centering
    \subfloat[]{\includegraphics[width=0.8\columnwidth]{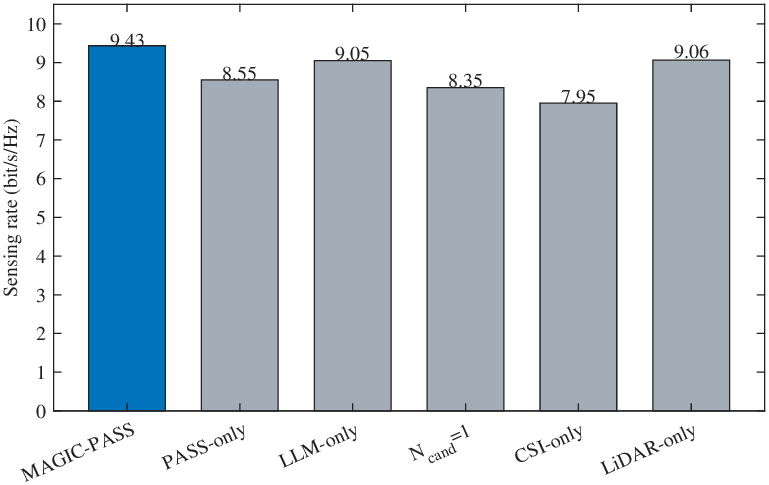}
    \label{ablation_sensing}}
    \hfill
    \subfloat[]{\includegraphics[width=0.8\columnwidth]{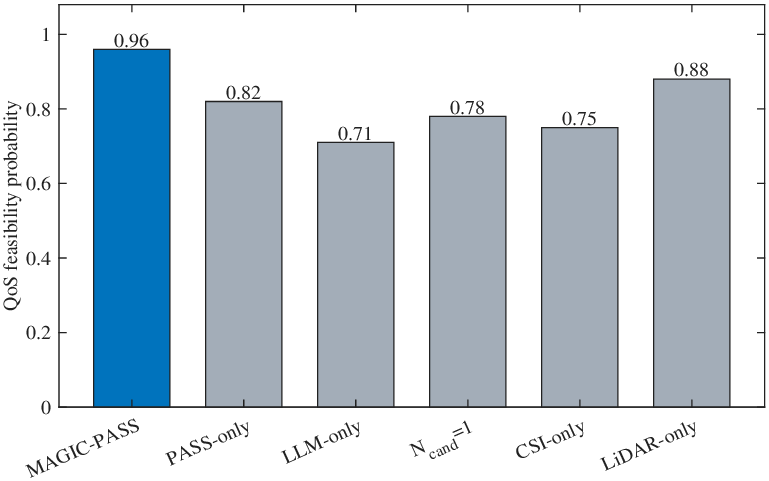}
    \label{ablation_qos}}
    \caption{Effects of the MAGIC-PASS components on (a) sensing rate and (b) QoS-feasibility probability.}
    \label{ablation}
\end{figure}

Fig.~\ref{ablation} evaluates the contributions of the configuration-generation and refinement modules and multimodal conditioning. The PASS-only variant removes all learned modules and employs the conventional PASS optimizer. The LLM-only variant retains multimodal tokenization, the frozen LLM, and the structured candidate decoder, but selects directly from the analytically mapped candidates without conditional diffusion, refinement, or feasibility correction. In these single-modality variants, the omitted observations are not supplied to the LLM. The user and target positions and blockage conditions remain available to the physical-guidance and evaluation procedures, as required to compute $\mathcal L_{\rm phy}$ and $\mathcal V_{\rm QoS}$. The $N_{\rm cand}=1$ variant changes only the number of candidate branches and retains the same diffusion, physics-guidance, and correction procedures as the complete method. The CSI-only and LiDAR-only variants retain the task token $\boldsymbol\tau$ and the complete generation-then-refinement procedure, while using only the indicated environmental modality. All variants use the same test scenarios, physical parameters, and feasibility tolerance.

Selecting directly from the LLM-generated configurations yields a lower sensing rate and QoS-feasibility probability than the complete MAGIC-PASS procedure. This comparison shows the benefit of refining the generated configurations and explicitly addressing their communication-QoS violations. MAGIC-PASS also performs better than the LiDAR-only and CSI-only variants under the same test scenarios, since the available environmental observations improve configuration generation. Moreover, the complete generation-then-refinement procedure outperforms direct selection from the LLM-generated configurations in both sensing rate and QoS-feasibility probability. Thus, generating candidate configurations before refining them improves the quality of the final PASS configuration.

The proposed MAGIC-PASS achieves the highest sensing rate of $9.43$~bit/s/Hz and the highest feasibility probability of $0.96$. The corresponding values for PASS-only are $8.55$~bit/s/Hz and $0.82$, such that MAGIC-PASS improves the sensing rate by $10.3\%$ and the feasibility probability by $14$ percentage points. LLM-only attains $9.05$~bit/s/Hz but only $0.71$ feasibility. This result also confirms that analytical parameterization guarantees the hardware constraints, whereas explicit physics-based refinement remains important for meeting the communication QoS constraints.


Using a single candidate reduces both the average sensing rate and the QoS-feasibility probability. Since the refinement and correction procedures are otherwise identical, this comparison shows the benefit of starting from multiple generated configurations. The LiDAR-only variant also outperforms the CSI-only variant in the considered setting. LiDAR can provide current three-dimensional scene geometry for blockage, and locations of users and target for candidate generation, whereas historical CSI records channels under previous PA configurations. Thus, multimodal environmental information can enhance the adaptability of MAGIC-PASS across user--target geometries and blockage states compared with single-modality variants. Both variants perform worse than the complete MAGIC-PASS, indicating that combining the available environmental observations improves configuration selection in the considered scenarios.

\section{Conclusion}
\label{Section: Conclusion}

This paper has proposed MAGIC-PASS for joint transmit-PA positioning, receive-PA positioning, and digital beamforming in PASS-enabled ISAC. A sensing-rate maximization problem has been formulated by jointly optimizing the transmit-PA positions, the single receive-PA position, and digital beamforming.
To address this nonconvex problem, we have developed offline-training and online-optimization procedures that generate multiple candidate PASS configurations and refine them through conditional diffusion and physical guidance. The PA-placement and power budget constraints have been enforced through the configuration mapping, while communication QoS has been checked before final selection. Numerical results have shown that MAGIC-PASS achieves higher sensing rates and QoS-feasibility probabilities than the conventional PASS- and MIMO-based benchmarks. Comparisons among different components of MAGIC-PASS have further shown the complementary benefits of multimodal conditioning, multi-candidate generation, and configuration refinement.


\bibliographystyle{IEEEtran}
\def\baselinestretch{1}
\bibliography{PASS}

\end{document}